\documentclass[onecolumn]{aa}

\usepackage{graphicx}
\usepackage{amsmath,bm}
\usepackage{txfonts}
\usepackage{hyperref}
\usepackage{float}

\graphicspath{{./}{figures/}}
\usepackage{xcolor}
\usepackage{soul}

\begin{document}
   \title{Response of the solar atmosphere to flux emergence}
   \subtitle{With emergence-driven prominence formation}
   \author{Xiaohong Li\inst{1}
          \and
         Yuhao Zhou\inst{2}
          \and
          Rony Keppens \inst{2}
          }
   \institute{Max Planck Institute for Solar System Research, Göttingen D-37077, Germany \\
              \email{lixiaohong@mps.mpg.de}
              \and
              Centre for mathematical plasma-astrophysics (CmPA), KU Leuven,
              Celestijnenlaan 200B, 3001 Leuven, Belgium
              }          
   \date{Received ; accepted }

  \abstract
   {Flux emergence is crucial for the formation of solar active regions and triggering of various eruptions. Observations show that solar activities including filament eruptions, jets, and flares are often associated with flux emergence. However, the detailed mechanisms by which flux emergence drives these eruptions remain unclear and require numerical investigation.}
   {Using 2.5-dimensional magnetohydrodynamic simulations, we investigate the interaction between emerging flux and background magnetic fields and dynamics of the induced eruptions.}
   {Our simulations model a stratified solar atmosphere, incorporating key energy transfer mechanisms such as radiative cooling, thermal conduction, and background heating. By systematically varying the strength and angle of the emerging magnetic field relative to the background field, we investigate its impact on the initiation and evolution of solar eruptions.}
   {This study extends our previous work, where a multi-thermal jet formed as emerging flux interacted with a pre-existing arcade hosting coronal rain. The simulations show that magnetic reconnection between the emerging flux and background field drives the formation of current sheets, magnetic islands and multi-thermal jets. Stronger magnetic fields result in earlier eruptions, more energetic jets, and enhanced heating. The formation and ejection of magnetic islands affect the structure and dynamics of the jet. When the hot and cool components of jets reach the other footpoint of magnetic loops, they will generate spicules near the transition region. Varying the angle between the emerging flux and the background field, we find that larger angles delay filament ascent and eruption timing but facilitate filament formation. Filaments form a hot shell and oscillate with a period of 10 minutes driven by periodic plasma ejections. Repetitive reconnection events inject cold plasma into the self-consistently formed filament channel, introducing a new prominence formation mechanism by flux-emergence-fed injection.}
   {Our analysis highlights the dynamic interplay between magnetic reconnection, plasma cooling and heating, and filament dynamics. These findings provide insights into solar eruptions and their observational signatures, emphasizing the role of multi-thermal structures in the corona.}
   \keywords{Solar atmosphere, Magnetohydrodynamical simulations, Solar activity, Solar prominences, Solar magnetic flux emergence}
   \maketitle

\section{Introduction} \label{sec:intro}

Magnetic flux emergence is widely recognized as a key driver of solar eruptive events, including jets, flares, and coronal mass ejections (CMEs) \citep{Zwaan1987, Shibata1989, Archontis2004}. The rise of magnetic flux from the convection zone into the solar corona is fundamental to the Sun's magnetic activity \citep{Pariat2009, Cheung2014}. Emerging magnetic flux can form active regions --- areas of intense magnetic activity responsible for solar eruptions such as flares and CMEs \citep{Fan2001}. When the newly emerging flux interacts with the preexisting coronal magnetic field, it often results in magnetic reconnection --- a reconfiguration of magnetic field lines that releases a large amount of energy \citep{Priest2002}. This energy release powers solar eruptions, including flares, jets, and even CMEs \citep{Inoue2018}. Magnetic reconnection plays an important role in driving eruptive solar phenomena and contributes to the formation of complex magnetic structures such as prominences and filaments, which are often precursors to solar eruptions \citep{Mackay2006, Schmieder2014, Syntelis2019}.

Given the observational challenges of studying flux emergence below the photosphere, numerical simulations have significantly advanced our understanding of these processes. Early work by \citet{Matsumoto1993} and \citet{Fan2001} explored the dynamics of magnetic flux tubes rising from the convection zone into the corona. Subsequent studies investigated how various factors, such as the twist, buoyancy, and strength of the flux tubes, influence the emergence process and magnetic reconnection \citep{Archontis2004}. Simulations demonstrated that the twist of a flux tube can enhance reconnection, leading to the formation of magnetic islands, a common feature observed during solar eruptions. Recent studies have introduced additional complexities, such as partial ionization \citep{Leake2013}, which influences reconnection rates and the formation of flux ropes in the upper atmosphere. \citet{Kusano2012} studied the onset of solar eruptions in different magnetic field structures, and proposed that magnetic helicity and details in the magnetic field perturbations are important for triggering flares and CMEs. Moreover, \citet{Leake2022} highlighted the critical role of the relative orientation between the emerging flux and the pre-existing coronal magnetic field, showing that it affects reconnection efficiency and the likelihood of large-scale eruptions.

Magnetic flux emergence has been employed in many simulations to trigger solar jets. \citet{Shibata1992} provided the first detailed model of coronal jets, driven by reconnection between emerging flux and ambient coronal magnetic fields. Their two-dimensional (2D) simulations illustrated how both hot and cool jets can be generated through this process, forming a framework for later observational and simulation-based studies \citep{Pariat2009, Wyper2017}. \citet{Yokoyama1995, Yokoyama1996} expanded this work by exploring the impact of different magnetic field orientations. It was shown that the relative angle between the emerging flux and the ambient field can significantly influence the jet's temperature, density, and overall dynamics \citep{Pariat2015}.

Advancements in computational resources have facilitated increasingly detailed three-dimensional (3D) simulations, leading to significant progress in understanding the dynamics of solar jets. \citet{MorenoInsertis2008} conducted one of the first 3D magnetohydrodynamic (MHD) simulations of flux emergence in a coronal hole, showing the formation of collimated, high-velocity jets that closely resemble observations from instruments like Hinode \citep{Cirtain2007}. Their work revealed that reconnection between emerging flux and the coronal magnetic field generates not only fast jets but also heated plasma loops \citep{Takasao2013}. Further studies by \citet{MorenoInsertis2013} highlighted the 3D structure of jets, which often take on a semi-hollow cylindrical shape, with fast outflows along the surface and slower, denser plasma in the central regions. These simulations also demonstrated the intermittent nature of jets, wherein repeated reconnection events give rise to both quiescent and explosive phases.

Recent simulation studies have further refined our understanding of jet formation, particularly regarding the influence of various physical parameters on jet properties. \citet{Archontis2013} investigated how the twist in emerging flux tubes interacts with the ambient coronal field, triggering so-called blow-out jets. These jets are characterized by their wide, filamentary structures and are driven by the untwisting of magnetic fields during the eruption. \citet{Lee2015} showed that torsional Alfvén waves, generated by reconnection between twisted emerging flux and untwisted pre-existing fields, play a critical role in driving collimated jets. Studies including those by \citet{Wyper2017}, confirmed that the reconnection-driven untwisting of magnetic fields is a key driver of fast, high-velocity jets.

Building on these developments, our previous work has conducted 2.5D MHD simulations to investigate the role of flux emergence in solar jet formation \citep{LiX2023}. These simulations successfully reproduced several key observational features, such as collimated jets and the formation of magnetic islands through magnetic reconnection between the emerging flux and the background coronal field. The tearing mode instability, often responsible for the formation of magnetic islands \citep{Ni2017}, is frequently observed during jet formation. For example, \citet{Singh2012} reported the formation of bright blobs in a current sheet-like region, exhibiting bidirectional motion --- both upward along the jet and downward toward the solar surface. These results are consistent with both observations and previous simulations, highlighting the role of reconnection and instability in solar jet dynamics \citep{Shibata2007, Raouafi2016}. A novel ingredient in our model was to start flux emergence in a coronal arcade that already demonstrated complex thermal non-equilibrium dynamics, with coronal rain in progress. In this work, we will emphasize especially this multi-thermal aspect in highly resolved 2.5D simulations, and we discover a new filament formation mechanism along the way.

As demonstrated earlier, physical parameters play a pivotal role in shaping the dynamics of jet formation in simulations. To further explore these dynamics, we now conduct a parameter survey to examine how variations in the angle between emerging flux and the background field, as well as changes in the strength of the emerging flux, influence jet shape and the formation of magnetic islands. The alignment angle governs the geometry and efficiency of magnetic reconnection, directly impacting jet morphology and velocity. Conversely, the strength of the emerging flux determines the amount of magnetic energy available for conversion into kinetic energy during the eruption. These geometry and energy parameters were chosen for this study because they represent fundamental aspects of flux emergence that dictate the onset and evolution of jets. Understanding their influence is essential for capturing the wide range of jet behavior observed in the solar corona. We aim to provide new insights into how different magnetic configurations influence the evolution of solar jets and their eruptive characteristics. This paper is structured as follows: Section \ref{sec:setup} describes the numerical setup. Section \ref{sec:result} presents the results of the parameter survey, focusing on the effects of the emerging magnetic flux's angle and strength on jet morphology and magnetic island formation. Finally, Section \ref{sec:conc} discusses the broader implications of this study.

\section{Numerical setup} \label{sec:setup}

The simulations here follow the setup in our work \citep{LiX2023}, and the parallelized Adaptive Mesh Refinement Versatile Advection Code \citep[\href{https://amrvac.org/}{MPI-AMRVAC},][]{Keppens2012, Porth2014, Xia2018, Keppens2021, Keppens2023} is employed. We use a Cartesian box covering $-60$ Mm $\le$ $x$ $\le$ 60 Mm and 0 Mm $\le$ $y$ $\le$ 60 Mm. An effective resolution of 39 km in both directions is achieved by using five AMR levels on a base grid of 192 $\times$ 96.

We include the effects of gravity, field-aligned thermal conduction, and optically thin radiative cooling, following the implementation described by \citet{LiX2022}. The set of MHD equations solved is:

\begin{equation}
\frac{\partial \rho}{\partial t} + \nabla \cdot (\rho \mathbf{v}) = 0,
\end{equation}
\begin{equation}
\frac{\partial (\rho \mathbf{v})}{\partial t} + \nabla \cdot \left( \rho \mathbf{v} \mathbf{v} + p_{\mathrm{tot}} \mathbf{I} - \frac{\mathbf{B} \mathbf{B}}{\mu_0} \right) = \rho \mathbf{g},
\end{equation}
\begin{equation}
\frac{\partial E}{\partial t} + \nabla \cdot \left[ \left( E + p_{\mathrm{tot}} \right) \mathbf{v} - \frac{(\mathbf{v} \cdot \mathbf{B}) \mathbf{B}}{\mu_0} \right] = \rho \mathbf{g} \cdot \mathbf{v} + \nabla \cdot (\boldsymbol{\kappa} \cdot \nabla T) - R + H,
\end{equation}
\begin{equation}
\frac{\partial \mathbf{B}}{\partial t} + \nabla \cdot (\mathbf{v} \mathbf{B} - \mathbf{B} \mathbf{v}) = 0,
\end{equation}
where $\mathbf{v}$ is the velocity, $\rho$ the mass density, $\mathbf{B}$ the magnetic field, and $\mathbf{I}$ the identity tensor. Assuming a fully ionized plasma with a hydrogen-to-helium abundance ratio of 10:1, the mass density is given by $\rho = 1.4\, m_{\mathrm{p}} n_{\mathrm{H}}$, where $n_{\mathrm{H}}$ is the hydrogen number density and $m_{\mathrm{p}}$ is the proton mass. The total pressure is defined as $p_{\mathrm{tot}} = p + B^2 / (2 \mu_0)$, comprising thermal and magnetic contributions. The thermal pressure $p$ is obtained from the ideal gas law: $p = 2.3\, n_{\mathrm{H}} k_{\mathrm{B}} T$. The gravitational acceleration is given by $\mathbf{g} = -g_{\odot} R_{\odot}^2 / (R_{\odot} + y)^2\, \hat{\mathbf{y}}$, where $R_{\odot}$ is the solar radius and $g_{\odot} = 274~\mathrm{m\,s^{-2}}$ is the gravitational acceleration at the solar surface. The total energy density is defined as $E = p / (\gamma - 1) + \rho v^2 / 2 + B^2 / (2 \mu_0)$, with adiabatic index $\gamma = 5/3$. Field-aligned thermal conduction is described by $\boldsymbol{\kappa} = \kappa_{\parallel} \mathbf{b} \mathbf{b}$, where $\mathbf{b} = \mathbf{B} / B$ is the unit vector along the magnetic field and $\kappa_{\parallel} = 10^{-6} T^{5/2}$~erg\,cm$^{-1}$\,s$^{-1}$\,K$^{-1}$ is the Spitzer conductivity. Radiative cooling is included via the term $R = 1.2\, n_{\mathrm{H}}^2 \Lambda(T)$, where $\Lambda(T)$ is the optically thin radiative loss function, interpolated from the tabulated values of \citet{Colgan2008}. For temperatures below $10^4$\,K, we set $\Lambda(T) = 0$ to account for the transition to optically thick and partially neutral plasma.

For the initial condition, a bipolar arcade loop system with a magnetic field strength of $B_0 = 20$~G is imposed, forming a constant angle of $30^\circ$ with the neutral line ($x = 0$, $y = 0$) and covering the entire domain. An exponentially decaying background heating balances the energy losses until the corona reaches a quasi-equilibrium state. Then we add a temporally and spatially randomly distributed heating at the footpoints of the loops to the energy input from the lower solar atmosphere, which is important when studying the details of the formation of prominences and coronal rain \citep{Zhou2020, LiX2022, LiX2023, Valeriia2023}. After we turn on the localized heating for a while, condensations start to grow continuously in the loop due to local runaway caused by thermal instability. At around $t_0$ = 144.6 minutes, the loops are filled with coronal rain, and then we put an emerging dipole at the left footpoint. The center of the dipole is initially located at the position ($x_0$, $y_c$), where $x_0 = -30$ Mm, $y_c = -10$ Mm. It is assumed to move upwards at a constant speed $\bm{v}$ = $v_0$ $\hat{\mathbf{y}}$ and $v_0$ = 2 km s$^{-1}$, then the center of the dipole is ($x_0$, $y_0(t)$) where $y_0(t) = y_c + v_0 \times (t - t_0)$. The emerging magnetic field can be written as 
\begin{equation} 
\begin{aligned}
B_{1x} &= B_1 \cos \left (\frac {\pi (x-x_0)}{L} \right ) \sin \phi  {\rm exp} \left (- \frac {\pi \sin \phi (y-y_0(t))}{L}\right ),  \\
B_{1y} &= - B_1 \sin \left (\frac {\pi (x-x_0)}{L} \right ) {\rm exp} \left (- \frac {\pi \sin \phi (y-y_0(t))}{L}\right ),   \label{equa} \\
B_{1z} &= B_1 \cos \left (\frac {\pi (x-x_0)}{L} \right ) \cos \phi  {\rm exp} \left (- \frac {\pi \sin \phi (y-y_0(t))}{L}\right ).
\end{aligned}
\end{equation}
Here $L$ = 20 Mm and $\phi$ is the angle between the loops and the neutral line (x = 0, y = 0). The emerging dipole is then introduced into the computational domain via time-varying electric field at the lower boundary \citep{FanY2004}: 
\begin{equation}
\bm{E}= - \bm{v} \times \bm{B_1}.
\end{equation}
After around 86 min ($t= $ 230.6 min), we stopped the artificial emergence and let the system evolve. By controlling the parameter $B_1$ and $\phi$, we can change the intensity of the emerging field and the angle between the emerging field and background field.

The simulation utilizes a three-step Runge-Kutta time integration technique, incorporating the ``van Leer" limiter and the Harten-Lax-van Leer Riemann solver. To maintain the magnetic field divergence at machine precision zero in one prechosen discretization, the upwind constrained transport (CT) method is strategically employed. The handling of field-aligned thermal conduction is managed using the Runge-Kutta Legendre super-time-stepping (RKL-STS) scheme \citep{Meyer2012, Meyer2014}, while the optically thin radiative losses are addressed through the exact integration method \citep{Townsend2009}. Additionally, the transition region adaptive conduction (TRAC) method \citep{Johnston2019, ZhouY2021} is integrated into the process to correct the underestimated enthalpy in the corona.

Regarding boundary conditions, the left and right boundaries are configured to have symmetrical density, total energy, and the $y$ component of momentum, along with $B_y$, while the $x$ and $z$ components of momentum, $B_x$ and $B_z$, are set to anti-symmetric. At the lower boundary, the velocity is uniformly zero, except within the emergence area where it is set to 2 km s$^{-1}$, as denoted by $v_0$. Here, density and pressure remain constant as per the initial conditions. At the upper boundary, a fixed zero velocity is maintained, and pressure and density are extrapolated in accordance with gravitational stratification. The values of $B_x$ and $B_z$ are determined through a zero-gradient extrapolation, and $B_y$ is derived using the zero-divergence constraint, applied at both the upper and lower boundaries.

\section{Results} \label{sec:result}
\subsection{Different emerging field intensity}
  
\begin{figure}[p]
  \centering
  \includegraphics[width=18cm]{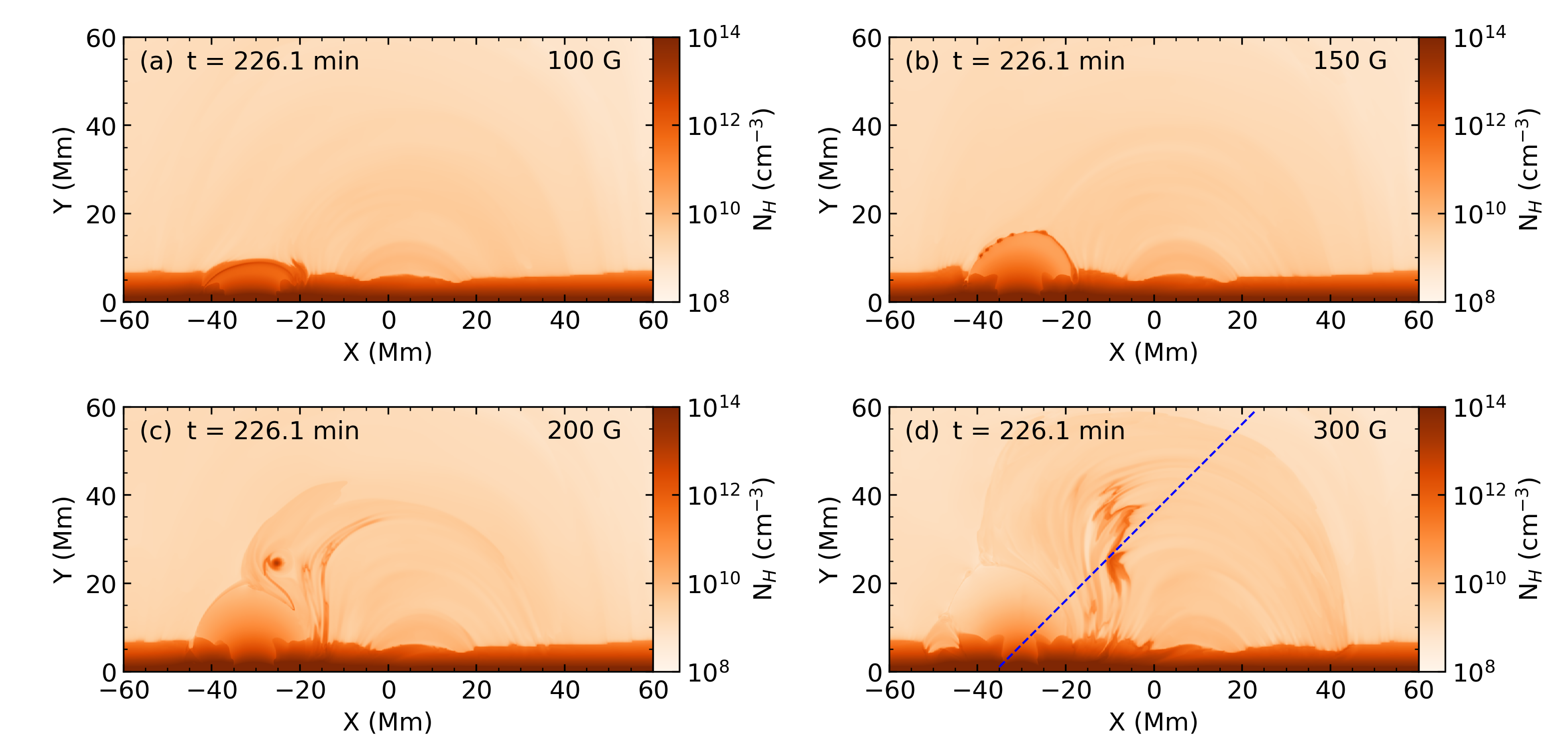}
     \caption{Number density maps showing the emergence with different field strength of B$_1$ = (a) 100 G, (b) 150 G, (c) 200 G and (d) 300 G. The dashed line in panel (d) indicates the slice position for Figure \ref{Fig2}. An animation of this figure (Movie1.mov) is available.}
        \label{Fig1}
  \end{figure}

\begin{figure}[p]
  \centering
   \includegraphics[trim=2cm 2cm 2cm 2cm, clip, width=\textwidth]{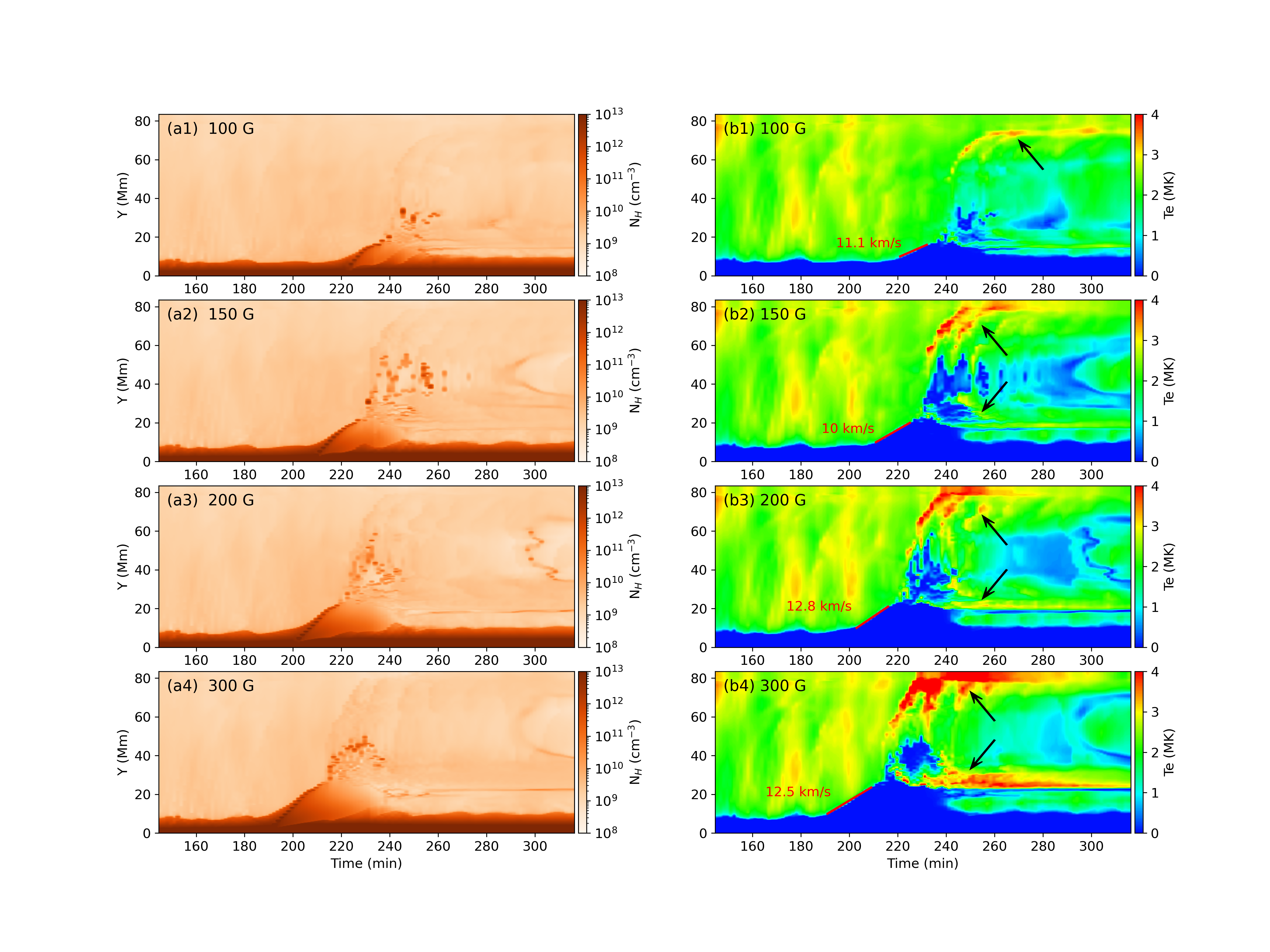}
     \caption{Time evolution of number density (panels (a1)$-$(a4)) and temperature (panels (b1)$-$(b4)) along the slice in Fig.\ref{Fig1}(d) with different emerging field strength. The black arrows in the right panels denote high temperature regions formed because of heating.
       }
        \label{Fig2}
  \end{figure}

As demonstrated in our previous study, the background magnetic field represents pre-existing coronal magnetic loops with $B_0$ = 20 G and a $30^\circ$ angle with respect to the neutral line ($x = 0$, $y = 0$). Then, a polarity-reversed bipolar field is added at the footpoint on the left side of the background loop, gradually extending into the corona. In this work, we investigate the effects of different emerging magnetic field strengths, specifically $B_1$ = 100 G, 150 G, 200 G, and 300 G, and $\phi$ = 0 for all cases. The overall evolution is demonstrated in the online supplementary Movie1.mov. We can see that the emerging magnetic fields of different intensities enter the corona at different times. Although we set the same emergence speed, the stronger the magnetic field intensity is, the greater the pressure it exerts on the coronal plasma, and the faster it enters the corona. At the boundary between the background field and the opposite-polarity emerging field, current sheets are formed, which can be observed in Movie1.mov. Over time, the current sheet elongates, accompanied by the formation of an arched filament in this region. As a result of the tearing instability, magnetic islands form within the current sheet. Subsequently, these islands are propelled upward, leading to the generation of multi-thermal jets. Figure \ref{Fig1} displays the density maps at the same moment $t = 226.1$ min under different emerging magnetic field strengths. It is evident that higher field strengths accelerate the overall evolution. For instance, in the 100 G case, the emerging loop has only just entered the corona (see panel (a)), while in the 300 G case, the jet has already formed (see panel (d)). To further investigate the evolution, we selected a slice originating from the emergence location and aligned it at a 45$^\circ$ angle. The position of the slice is indicated by a dashed line in panel (d). The density and temperature evolutions along this slice are presented in Figure \ref{Fig2}. From top to bottom, the panels correspond to $B_1$ values of 100 G, 150 G, 200 G, and 300 G. We find that as the magnetic field strength increases, the moment the emerging field enters the corona is earlier (from about 220 min to around 190 min). The arched filament formed near the current sheet can be distinguished by the dense front, and its rising is marked by the red lines in the temperature plot. The emerging speed of the arched filament is approximately 10 km s$^{-1}$ in all cases, and this kind of filament ascent has been observed and recognized as a pre-eruption phase of solar flares or other eruptions \citep{Sterling2005, McCauley2015}. Subsequently, magnetic islands form and erupt, with eruption times of 240 min, 230 min, 221.8 min, and 214.7 min respectively. 

\begin{figure}[p]
  \centering
    \includegraphics[trim=0cm 2cm 0cm 2cm, clip, width=\textwidth]{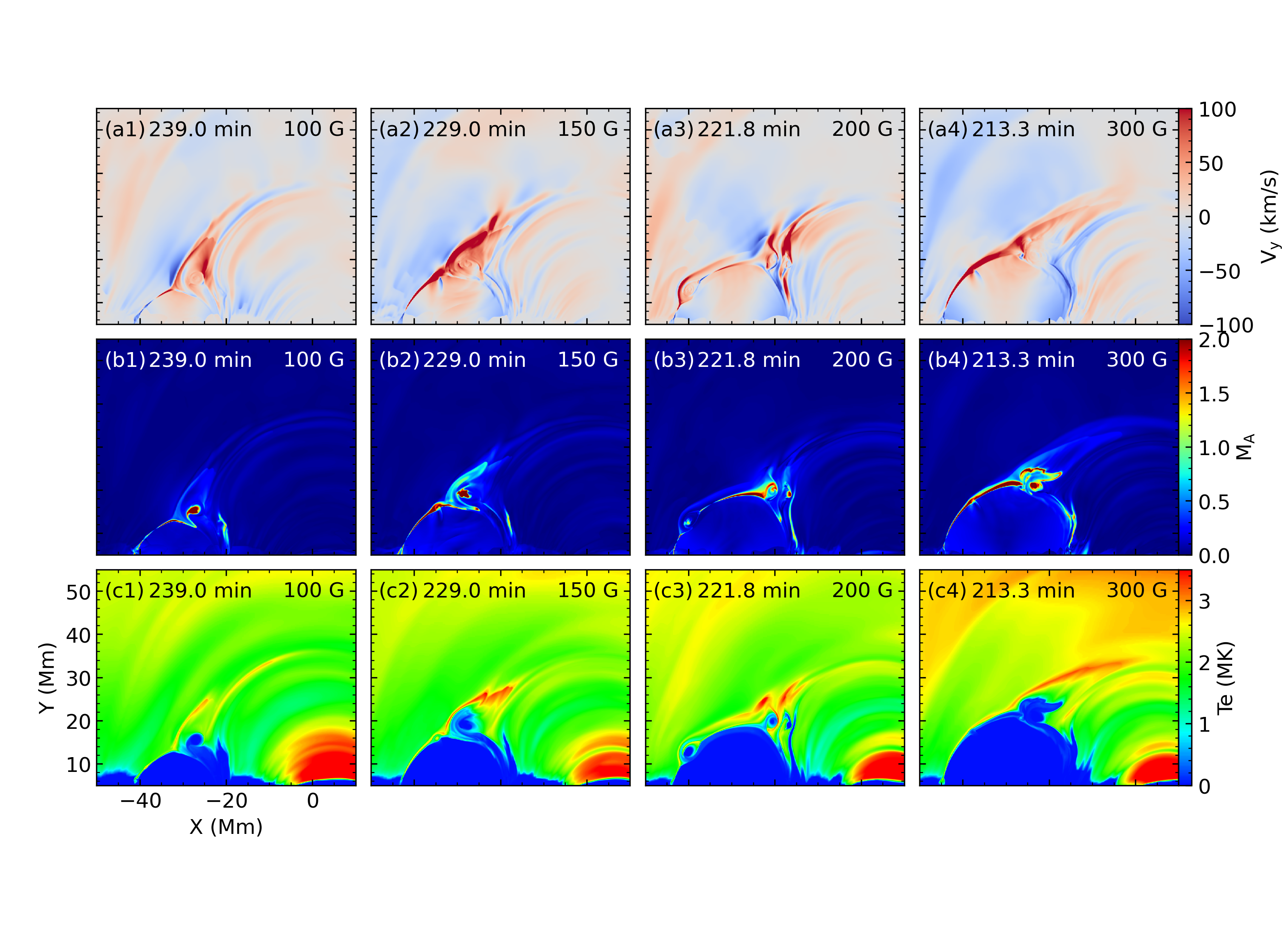}
     \caption{Vertical velocity component (panels (a1)$-$(a4)), Alfvén Mach number (panels (b1)$-$(b4)), and temperature (panels (c1)$-$(c4)) maps displaying the magnetic islands and heating with different emerging field strength at different time.
       }
        \label{Fig3}
  \end{figure}

As noted earlier, magnetic islands are formed within the current sheet due to the reconnection between the emerging magnetic field and the original magnetic field. Due to coalescence instability \citep{FinnKaw1977}, these magnetic islands merge to form large plasma blobs. Figure \ref{Fig3} shows the $y$-direction velocity, Alfvén Mach number $M_A$ and temperature maps of the formed large blobs from top to bottom panels. As shown in Figure \ref{Fig3} , we can see that under conditions of $B_1$ = 100 G, 150 G and 300 G, a large plasma blob is formed, primarily located above the arch-like structure. Whereas, in the case of $B_1$ = 200 G, in addition to a large plasma clump located at the right apex, another large magnetic island is also formed on the left side of the arch-like structure (see panel (c3)). As part of the magnetic reconnection process, the plasma blobs are ejected out (see panels (a1)$-$(a4)). This is likely facilitated by the melon seed ejection mechanism, as described by \citet{Kippenhahn1957}. This mechanism activates when an imbalance occurs in the Lorentz force that stabilizes the plasmoids along the sides of the current sheet. The ejected velocity is in the order of the local Alfvén velocity, as $M_A$ near the blobs is around or more than one. In front (the upper-right region) of these plasma clumps, significant heating occurs, reaching temperatures of several million Kelvin (MK), forming hot jets as shown in Figure \ref{Fig3} (c1)$-$(c4). The heating patterns align closely with the areas of high pressure. This heating effect is not limited to the front of the plasma blobs; it also extends from the current sheet to the arch-like structure below. The temperature slice images in Figure \ref{Fig2} reveal that higher magnetic field strengths correlate with more noticeable heating effects and higher peak temperatures. As indicated by the black arrows in the right panels of Figure 2, the heating in front of the plasma blobs and down to the arch structures could both be clearly observed. Additionally, the heating follows a certain periodic pattern, which becomes more apparent at higher magnetic field strengths. For example, there are two obvious periods in Fig. \ref{Fig2}(b2), and four periods in Fig. \ref{Fig2}(b4). The period of the heating is around 10 minutes. As mentioned in our previous paper, the jets that form due to this ejection process exhibit whip-like movements, resulting in periodic oscillations. 

\begin{figure}[p]
  \centering
   \includegraphics[trim=4cm 0cm 3cm 0cm, clip, width=\textwidth]{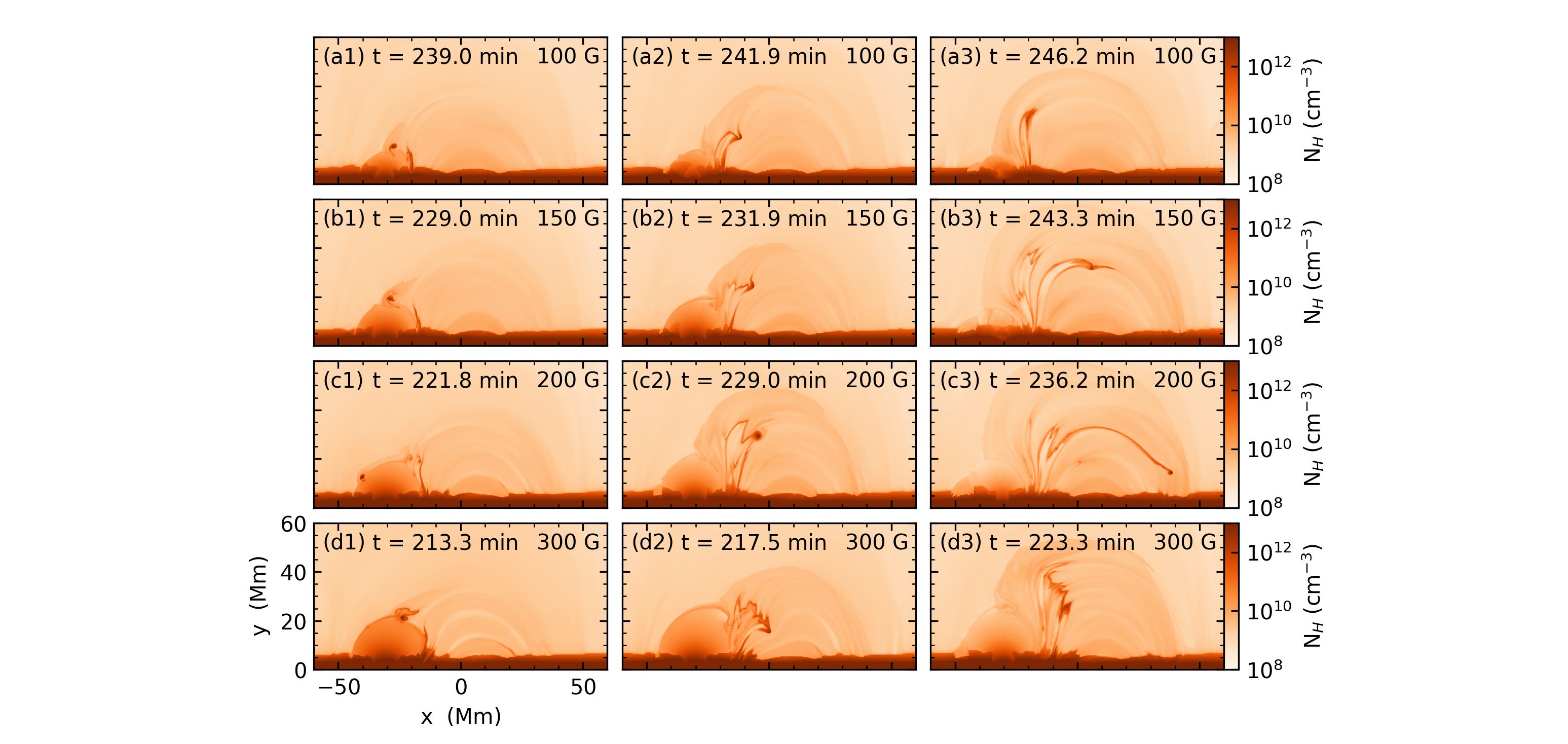}
     \caption{Number density maps showing the eruption with different emerging field strength. The times (from left to right) for the 4 field strength cases (top to bottom) are chosen such that they represent similar stages in the overall jet eruption scenario.
       }
        \label{Fig4}
  \end{figure}

After the magnetic islands coalesce into larger structures, these cold plasma blobs --- composed of relatively low-temperature material from the lower solar atmosphere that has not been fully heated or ionized --- are ejected outward, forming a jet with a temperature of about 6000 K. Figure \ref{Fig4} illustrates the ejection of the plasma blobs. For magnetic field strengths of 100, 150, and 300 G, the jets ascend to a certain height but fail to reach the opposing magnetic footpoint; instead, they fall back to the location where they were ejected from. Only in the scenario with a 200 G magnetic field does the jet material successfully reach the footpoint on the other side of the magnetic loop. Our analysis indicates that this behavior is closely tied to the dynamics of magnetic island ejections. In the first three cases, a single plasma blob forms, and its ejection subsequently bends the existing magnetic loop on its right side, where the magnetic tension slows the jet, causing it to fall back. For the 200 G case, the plasma blob at the apex of the emerging magnetic loop on the right is also obstructed. However, the magnetic island situated on the left side of the loop continues its trajectory along the magnetic field lines, ultimately reaching the footpoint on the opposite side, as demonstrated in Movie1.mov. Overall, the morphology of the jet is intricately linked to the configuration of the magnetic field. Additionally, it is influenced by the number of magnetic islands and associated ejections. Within this context, both tearing and coalescence instabilities play crucial roles in shaping the observed dynamics.

\begin{figure}[p]
  \centering
   \includegraphics[trim=1cm 0.5cm 0cm 0cm, clip, width=\textwidth]{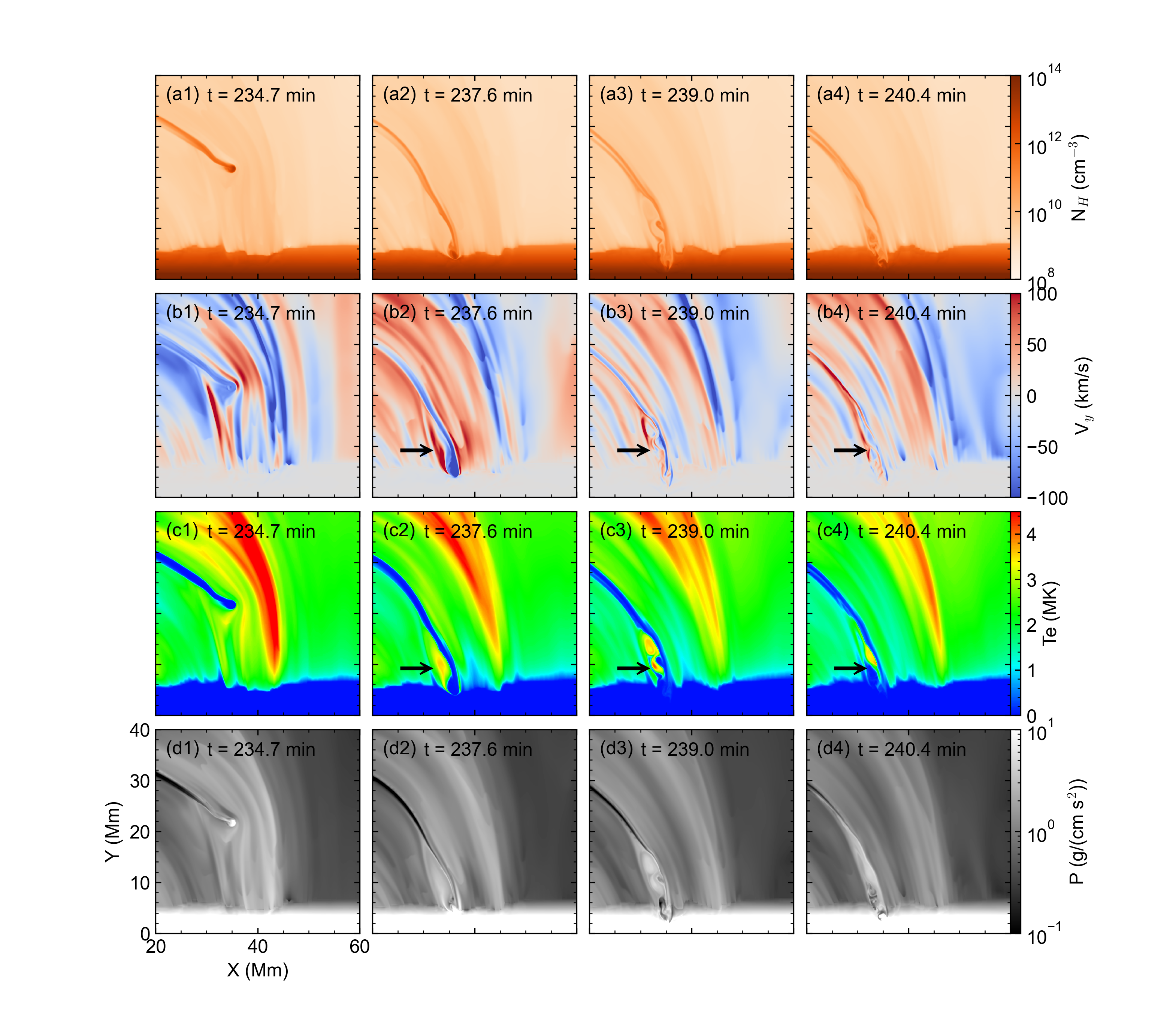}
     \caption{Number density, $y$-direction velocity, temperature, and pressure maps displaying the response of chromosphere and TR when the cool jet enters the lower solar atmosphere during case B$_1$= 200 G. The black arrows in the figure indicate the backflow.
    }
        \label{Fig5}
  \end{figure}

In the 200 G case, when the cool plasma jet reaches the footpoint on the opposite side, it triggers further reactions in both the transition region (TR) and the chromosphere, as depicted in Figure \ref{Fig5}. The jet moves at an approximate speed of 70 km s$^{-1}$. At the jet's head, the pressure builds up significantly, as shown in panels (d1)$-$(d4). Upon reaching the chromosphere, this pressure increase leads to localized compression and a subsequent backflow of material indicated by the arrows in panels (b2) and (c2). The velocity and temperature diagrams reveal that this backflow propagates at approximately 50 km s$^{-1}$, a phenomenon commonly associated with chromospheric shock interactions \citep{Martinez2017}. The process also induces significant localized heating, with temperatures rising up to 4 MK. Furthermore, at the interface between the backflow and the erupting material, small vortex-like structures are observed. These structures may represent manifestations of the Kelvin-Helmholtz instability (KHI), which is often seen in shear-flow environments in solar and astrophysical plasmas \citep[e.g.,][]{LiX2023}. Such instabilities could play a pivotal role in further mixing and energy dissipation within the system.

Since the jets are multi-thermal, prior to the cool component, the hot component also reaches the opposing footpoint and influences the TR and the chromosphere. The temperature plots in Figure \ref{Fig6}(a1)$-$(a4) reveal what happens when the hot jets arrive at the opposite footpoint in the case of $B_1 = 200$ G. Accompanying this hot jet movement, a notable backflow near the footpoint is evident in the $y$-direction velocity maps. This backflow significantly increases the density and pressure within the magnetic loop, leading to a rise in the plasma $\beta$ value, as shown in panels (b1) to (b4). This phenomenon is further visualized through both side-on and line-of-sight perspectives in H$\alpha$ imaging, presented in panels (d1) to (d4) and (e1) to (e4), respectively. The H$\alpha$ emission is calculated using the approximate method from \citet{Heinzel2015}. In the line-of-sight view, the emission below the TR is cut off and treated as background emission; see \citet{Jenkins2021} for details. In the side-on view, since this approximate method is designed for solar prominences/filaments in the corona, the emission below the TR is not reliable. Within these images, a pronounced brightening in the H$\alpha$ line aligns with the areas of backflow. In several instances of jet observations, an enhanced brightness at one footpoint is often detected in the 304 {\AA} or H$\alpha$ spectral bands before the jet material reaches its peak. This brightness increase is linked to the visibility of cooler chromospheric matter in these spectral ranges. Notably, the backflow observed here differs from the mechanism shown in Figure \ref{Fig5}, where backflow primarily arises from the ejection of cooler chromospheric material. In Figure \ref{Fig6}, by contrast, there is an initial heating phase in the loop before cooler plasma arrives, suggesting that the observed brightening in these spectral ranges results directly from heating processes in the loop rather than from ejected cool plasma. This points to a more complex interplay among the multiple thermal components of the jets and reveals how both hot and cool plasma dynamics can drive varying impacts on solar atmospheric structures.
  
For the cold component, it is noted that only in the scenario with a 200 G magnetic field does the cold matter reach the other side of the footpoint, triggering a backflow. However, the hot component, present across all magnetic field strengths ($B_1 = 100$, 150, 200, and 300 G), consistently produces brightenings in the H$\alpha$ passband, as shown in Figure \ref{Fig7}. Figure \ref{Fig7} displays only side-on and line-of-sight H$\alpha$ maps. In each case, the hot jet component successfully reaches the opposite side of the solar loops. These observations reveal backflows and H$\alpha$ brightenings at the footpoints for all tested magnetic field strengths. The brightenings are attributed to shock-induced heating and compression as the hot jet impacts the chromosphere and TR, consistent with findings in similar jet studies. The 200 G case, in particular, exhibits the most pronounced brightening. This enhanced emission is hypothesized to result from an increased pressure at the loop footpoint. Such interactions between the hot and cold jet components emphasize the role of thermal and magnetic forces in shaping jet dynamics.
  
To further investigate the origins of cold and hot plasma and the energy transfer dynamics under varying strengths of emerging magnetic fields, we consider all cells above 7 Mm in height as the region of solar corona. Within this coronal region, the plasma is categorized into three groups: “cool” plasma (\(T < 0.02\) MK), warm plasma (\(0.02\) MK \(- 0.5\) MK), and hot plasma (\(T > 0.5\) MK). Using the density, velocity, temperature, and magnetic field strength of each cell, we can calculate the mass of each temperature category, as well as the magnetic, kinetic, and internal energies at each snapshot. The temporal evolutions of mass, magnetic energy, kinetic energy, and internal energy are shown in Figure \ref{Fig8}. 
The intervals of forced emergence are marked with dashed lines in these panels. From panels (a) and (b) we can see that with increasing magnetic field strength, the rise in the mass of cold and warm matter occurs earlier, with cold plasma reaching higher mass values. This is attributed to the fact that higher magnetic field strengths trigger earlier eruptions and matter ejections, thereby accelerating the mass accumulation of cold and warm matter. Panels (d)$-$(f) suggest that stronger magnetic fields also increase the energy release into the corona, leading to more vigorous eruptions and higher kinetic energy in the ejected matter. Intriguingly, at 200 G, a second, significantly larger peak in the mass of warm and hot matter is observed (see the red curves in panels (b) and (c)), due to matter being ejected to the opposite footpoint at this particular strength, since the timing of this peak coincides with the ejection of cold matter to the other footpoint. This behavior supports previous simulations and studies indicating that matter reaching the other footpoint can trigger a backflow, resulting in systematic heating and an increase in the mass of warm and hot matter \citep{LiX2022}.

We also discuss here the longer-term evolution of the emerged loop system, beyond the times when jets form and interact. Subsequent to eruptions of the jets, we stop the emergence of flux at $t =$ 230.6 min, and consistent magnetic reconnection is still observed between the emerged magnetic loop and the pre-existing background magnetic field. As shown in Figure \ref{Fig9}, this reconnection process results in two regions with new magnetic field configurations. The blue curves in panels (a1) and (a3) indicate the magnetic field lines. At the thin upper edge of the current sheet, some field lines become open, propelling plasma outward, while closed coronal loops form to the left of the emerging dipole. As reconnection goes on, a structure with two parts of loops develops beneath the jet. The right section (emerged chamber) retains its original loop connections, while the left section (reconnected chamber) evolved into a new set of closed, high-temperature loops. Over time, the emerged chamber contracts laterally, whereas the reconnected chamber expands to encompass an area comparable to the initial size of the emerged chamber. Within the reconnection paradigm, cold matter is seen to be translocated from the emerged loop to a newly formed loop. The Atmospheric Imaging Assembly \citep[AIA;][]{Lemen2012} 94, 171 and 193 {\AA} synthetic images in Fig. \ref{Fig9} are created using the technique described in \citet{Xia2014}. Initial emergence fills the loop with cold, dense matter, as shown in the left panels. In the 94 {\AA} and 171{\AA} images, the bright structure corresponds to the emerging loop adjacent to the open magnetic field lines that have not yet reconnected. The ensuing reconnection phase produces a current sheet loaded with similarly cold, dense matter, visible in the middle panels as denoted by the red arrows. The current sheet is predominantly evident in the 94 {\AA}, with manifestations also in the 171 and 193 {\AA}. This phase also results in significant heating of the reconnected loops. Subsequent observations show that cold matter is transferred to the apex of the newly-formed loop, which subsequently undergoes contraction, channeling matter towards its foot point. In a recent work on a coronal cloud filament above a null point, a similar reconnection-aided mechanism of filament mass loss was demonstrated \citep{Craig2025}. The newly-formed loop is clearly visible in the 94 and 171 {\AA} passbands, located to the left of the original emerged loop. The red arrows in the right panels delineate the cold matter in the newly-formed loop. The transfer of cold plasma through the current sheet, as observed here, has been reported in prior studies of filament formation \citep[e.g.,][]{ZhaoXZ2022}, where it was instead acting to feed cold plasma into an erupting prominence. Overall, the observed transfer of cold material during reconnection and the subsequent evolution of loop structures underscore the dynamic interplay between plasma cooling, reconnection-driven heating, and magnetic field restructuring in the solar corona.
  
\subsection{Different emerging field angle}

As a second parametric study, we systematically varied the angle $\phi$ in Equation \ref{equa} by setting it to 0$^\circ$, 30$^\circ$ and 60$^\circ$, while fixing $B_1$ at 150 G. Since the background magnetic field is inclined at $30^\circ$ to the neutral line ($x = 0$, $y = 0$), and the polarity orientation of the background and emerging fields are opposite by construction, the relative angle between the two magnetic systems becomes $\phi + 30^\circ$. The adjustment modifies the angle between the emerging flux and the existing background magnetic field, resulting in angles of 30$^\circ$, 60$^\circ$, 90$^\circ$, respectively. In the following text, we use the relative angle instead of $\phi$ to describe the settings. As demonstrated in Figure \ref{Fig10} and the accompanying movie2.mov, although the overall morphology of the solar eruptions remains consistent across these different angular configurations, notable variations are observed in their scale and timing. A sequence of events is consistently observed, including the formation of an arched filament, the development of a current sheet, and the ejection of magnetic islands. The arched filament, a product of magnetic flux emergence, is first evident in the $30^\circ$ case, approximately at $t = 206.1$ min. As relative angle increases, the ascent of the filament into the corona is delayed, reaching lower atmospheric heights, and consequently, the timing of the eruption events is also postponed. Moreover, the magnetic field topologies after the eruption are significantly different. The boundary-driven emergence is stopped at $t=$ 229.0 min for all the cases. In the $30^\circ$ configuration, the emerging loops stabilize at their current altitude and continue connecting with the nearby background fields. For the $60^\circ$ case, the emerging loops reconnect with the background fields and continue to ascend slowly into the higher atmosphere. Remarkably, in the $90^\circ$ case, the emerging loops rise up to higher altitudes more rapidly, accompanied by filament formation due to mass ejection from the lower atmosphere.

We analyze the formation of the filament in the $90^\circ$ case in detail. The number density maps in Figure \ref{Fig11} illustrate the post-eruption phase, during which the emerging loops engage in continuous reconnection with the pre-existing magnetic field. This interaction causes the loops to stretch, become unstable, and expand towards the upper left, as shown in panels (a1) and (a2). Subsequently, reconnection happens between the legs of these stretched loops, leading to the formation of a filament channel, and cold plasma is injected into this filament channel through the reconnection process, as evident in panels (a4) to (a6). Continuous and repetitive reconnection results in the accumulation of cold mass within the filament channel, thereby elevating the filament's altitude while increasing its density and size. The arrow in panel (a7) highlights the cold plasma. From panel (a7) to (a9), we can see that the height of the filament increases from 22 Mm to around 40 Mm, and meanwhile the diameter of the filament changes approximately from 6 Mm to 10 Mm.

Figure \ref{Fig12} presents maps of the $y$-direction velocity and temperature, providing a comprehensive view of the dynamics and thermodynamics of the system. The emerging loop is characterized by a cooler temperature and lower pressure compared to its surroundings. The ejected plasma has a temperature similar to the chromosphere, and as it is ejected upwards, it develops a hotter shell (see panel (b2)). The black arrows in the temperature maps (b3) and (b5) indicate the presence of a hot current sheet. In the corresponding velocity maps, there are bi-directional flows on both sides of the current sheet, which provide evidence that reconnection happens between the legs of stretched loops. The upward ejection of plasma compresses the atmosphere ahead of it, leading to heating due to the resulting rebound shock, which is clearly depicted in panel (b5). This heating, in conjunction with the heating caused by the reconnection, contributes to the formation of a hot shell encasing the filament. As more and more plasma is ejected and forms a heavier filament, the shell surrounding the filament becomes progressively hotter and thicker. Additionally, the entire filament exhibits a rotational motion due to the ejection of lower plasma and the constraints imposed by the overlying magnetic field, as observed in panel (a6).
  
The filament shows an oscillation during its formation. To study it in more detail, we make a slice along the dashed line in panel (a4) from Figure \ref{Fig11}, and the time evolution of density, temperature and pressure along this position are shown in Figure \ref{Fig13}. In panel (a), dense plasma can be seen rising due to the emergence and forms a dense filament. Due to the plasma ejection from the lower atmosphere, the filament goes up and starts to oscillate. Each ejection raises the filament higher, with the amplitude being largest initially and gradually decreasing over time, consistent with previously observed filament oscillations \citep{Luna2014, Ni2022}. In the temperature map in panel (b), the filament is getting cooler and cooler, and each ejection generates heating in the front, forming a hot shell outside the filament together with the heating by the current sheet. Panel (c) shows the pressure variation along the slice, we can see that the hot shell corresponds to the high pressure area, the ejection presses the atmosphere ahead, and the filament forms and oscillates. The dashed lines in panel (c), derived from the number density map in panel (a), outline the filament core where the number density is the highest, revealing that the oscillation has a period of approximately 10 minutes, which aligns with typical oscillatory behavior in filament dynamics \citep{Tripathi2009, Arregui2018, Smirnova2022}. However, we now demonstrate for the first time that filament formation and growth, repetitive reconnection, and filament oscillations can be closely linked in specific combinations of flux emergence and background field topologies.
  
From the above analysis, it can be concluded that the formation and oscillation of the filament are multi-thermal processes, resulting in distinct manifestations across different AIA channels. We generated synthetic AIA 94, 193 and 304 {\AA} images, as shown in Figure \ref{Fig14}. An online movie (Movie3.mov) is also provided, showing the evolution of the entire region in AIA 193 {\AA}. Due to magnetic flux emergence, a cooler cavity formed within the nearly million-degree corona, as evident in the 94 {\AA} image in panel (a1). In other temperature-sensitive channels, chromospheric material within the ejection can be observed, as shown in panel (c1). After the material is ejected, the filament becomes visible in the 304 {\AA} channel, while the formed hot shell is observable in the 94 {\AA} channel. As the material continues to be ejected, the filament is pushed progressively higher. Due to Rayleigh$-$Taylor instability (RTI) and KHI, finger-like structures develop underneath the filament, as shown in the right zoomed-in panels. These instabilities are well-documented in plasma dynamics and play a significant role in filament formation and evolution \citep{Hillier2018, Jenkins2022, Changmai2023, Thomas2024}.

\section{Conclusion and Discussions}  \label{sec:conc}

This study presents an in-depth investigation of solar eruptions driven by the interaction between emerging magnetic flux and pre-existing coronal magnetic fields, using 2.5D MHD simulations. By varying the strength and orientation of the emerging flux, we elucidate the processes governing the formation, dynamics, and thermal evolution of jets and filaments in the solar atmosphere. 

We first explore the effects of different magnetic field strengths ($B_1$ = 100, 150, 200, and 300 G) on the dynamics of emerging solar magnetic loops. A reversed polarity bipolar field is forced into the atmosphere by a boundary-driven emergence, forming current sheets that drive magnetic reconnection. Stronger fields enter the corona faster, accelerating the evolution of arched filaments, current sheets, and magnetic islands formed by tearing instability. These islands coalesce into plasma blobs, propelling multi-thermal jets. The velocity of the jets is of the same order as the local Alfvén speed, consistent with reconnection model \citep{Priest2000}. The detailed dynamics of the jets vary with different magnetic field strengths. In the 100, 150, and 300 G cases, the jets fail to reach the opposing footpoint, falling back due to magnetic tension. In the 200 G scenario, plasma blobs successfully cross the loop, inducing significant heating and backflow in the chromosphere, with velocities up to 50 km/s and temperatures of 4 MK. Hot jets consistently reach the opposite footpoint, generating H$\alpha$ brightenings as argued by approximate spectroscopic synthesis across all cases. The 200 G case uniquely exhibits cold plasma triggering enhanced heating at the opposite footpoint. Post-emergence, reconnection reconfigures magnetic loops, transferring cold plasma into newly formed loops. Stronger fields increase energy release and plasma eruptions, highlighting the critical role of magnetic reconnection in shaping solar jets, plasma dynamics, and coronal heating. 

We also examine the effects of varying the angle between emerging magnetic flux and the background field ($30^\circ$, $60^\circ$ and $90^\circ$) on solar eruptions. Across all configurations, key events include current sheet formation, magnetic island development and jet eruptions. Smaller angles result in earlier jet eruption, while larger angles delay emergence but enable higher altitudes and filament formation via mass ejection. Here we only see the formation of a filament in the $90^\circ$ case, probably because when the emerging flux is more perpendicular (or forms a larger angle) with the background field, the opposing magnetic forces are stronger, potentially leading to more efficient magnetic reconnection. Also, the confinement from the overlying background field is less, which can result in chromospheric material escaping into the corona. In the $90^\circ$ case, continuous reconnection between the emerging flux and background field forms a filament channel. Cold plasma accumulates within this channel, increasing the filament's density, size, and altitude. A hot shell encases the filament, growing thicker as mass accumulates. The filament exhibits rotational motion due to plasma ejection and magnetic constraints. Filament oscillations with a 10-minute period accompany its formation, driven by repeated plasma ejections and atmospheric compression. Multi-thermal dynamics is manifest in synthetic AIA images: the filament is visible in the cooler 304 {\AA} band, while the hot shell appears in the 94 {\AA} and 171 {\AA} bands. These results highlight the interplay of reconnection, plasma dynamics, and instabilities in filament formation and oscillations. We propose that when the emerging flux is nearly anti-parallel to the background field, magnetic reconnection tends to be more direct and efficient. The released energy is more likely to convert into kinetic energy, making it easier to trigger jets. Meanwhile, when the emerging flux is closer to being perpendicular to the background field, while reconnection can still occur, the evolution is more likely to result in flux rope formation. A flux rope may then rise into the corona and favor the formation of an injection-based filament.

\subsection{Plasmoids}

The formation and dynamics of plasmoids are central to understanding the magnetic reconnection processes driving solar eruptions such as jets and filaments. Observational studies have demonstrated that plasmoids appear across a wide range of temperatures, from the chromosphere to the corona, showcasing their multi-thermal nature and diverse dynamic behaviors \citep[e.g.,][]{Singh2012, Kumar2019, Kumar2023, Mulay2023}. In our simulations, we also observe multi-thermal plasmoids during the formation of the arched filament and the eruption of jets. The morphology of the jet is intricately linked to the interplay of magnetic islands and associated ejections. Within this dynamical framework, tearing instability \citep{Priest2002} and coalescence instability \citep{Bellan2006} play pivotal roles. Our simulations reproduce key aspects of plasmoid generation through magnetic reconnection, consistent with previous studies \citep{Yang2013, Zhang2014}. Here, we also find that the ejection of plasmoids can contribute to the formation of prominence, similar to the mechanism found in \citet{ZhaoXZ2022} where the plasmoids feed the flux rope with chromospheric matter through the current sheet and lead to the formation of a prominence. Additionally, the oscillatory behavior we observe, with a period of approximately 10 minutes driven by periodic plasma ejections, suggests repetitive reconnection events at the current sheet, a feature also inferred from recurrent jets \citep{Yang2023}. These findings not only enrich the understanding of plasmoid behavior in solar jets and filaments, but also bridge observational insights and theoretical models. It is important to extend these studies to 3D simulations in the future to explore the full spatial complexity of plasmoid dynamics, including their 3D flux rope manifestations, and further investigate their role in solar magnetic reconnection processes.

\subsection{Instabilities}
In our simulations, we observed vortex-like or finger-like structures at different locations and stages, which are considered as manifestations of RTI and KHI. RTI and KHI are fundamental to understanding the dynamic evolution of solar plasmas. The KHI emerges at shear-flow interfaces where velocity differences exist between adjacent plasma layers. In our previous study \citep{LiX2023}, we have shown that KHI developed at the boundary of a jet when the downward plasma has velocity shear with the surrounding corona. In this study, we also observed vortex-like structures when the jet reached the footpoint and caused a backflow, as shown in Fig. \ref{Fig5}, which could similarly be attributed to KHI. Observationally, resolving KH vortices remains challenging due to the spatial resolution limits of current space- and ground-based instruments. So far, KHI has been observed in large-scale structures such as streamers \citep{Feng2013}, CMEs \citep[e.g.,][]{Foullon2011, Paouris2024} or through high-resolution observations \citep[e.g.,][]{LiX2018, Hillier2018a, Antolin2018, Yang2018}. However, it is likely that KHI is ubiquitous in the solar atmosphere, awaiting further exploration with improved instrumentation.

The RTI arises at the interface between a denser plasma overlying a lighter plasma in the presence of gravity or an effective acceleration. In our results, finger-like structures develop beneath filaments during their ascent, particularly in the $90^\circ$ flux emergence case. These features are characteristic of RTI, driven by the deceleration of the filament as it interacts with the surrounding corona. The efficient injection of cold plasma via reconnection enhances the density contrast, creating ideal conditions for RTI growth. Such instability not only shapes the lower boundary of the filament but also contributes to mixing between the cool filamentary material and the hotter coronal plasma \citep{Jenkins2022}. However, true RTI development will almost surely manifest itself more in actual 3D settings, as the current setup and its 2.5D nature in effect suppresses its growth. Observationally, \citet{Berger2017} detected RTI-like structures in prominence bubbles, providing indirect evidence of RTI in filament dynamics. These features are often associated with dense material launched upwards, which later falls under gravity, triggering the onset of RTI \citep{Hillier2018}. Smaller-scale phenomena, such as spicules and X-Ray/EUV jets, may exhibit RTI but remain beyond the current observational resolution. We note that RTI is mostly associated with quiescent filaments, such as simulated in \citet{Dion2024}, where again the 3D aspect proved key to allow for effective magnetic interchange. Here, the setup is targeting more active region filament formation, and our study suggests that small-scale dynamics due to RTI/KHI is clearly possible in that environment as well.

The simultaneous occurrence of RTI and KHI in our simulations highlights their interdependence in dynamic solar environments. The RTI-driven mass redistribution creates localized density and velocity gradients, which in turn seed KHI at filament boundaries. This interplay can drive turbulence in the filament environment, further promoting plasma mixing and heating \citep{Changmai2023}. Observationally, these instabilities may manifest as fine-scale structures and dynamic features in high-resolution solar images, particularly in the 304 {\AA} and 171 {\AA} channels. As observational capabilities improve, especially with the Solar Orbiter \citep{Muller2020}, the Daniel K. Inouye Solar Telescope \citep[DKIST,][]{Rimmele2020} and other next-generation instruments, these findings could provide deeper insights into the physical processes driving solar eruptions and their observational signatures.

\subsection{Prominence Formation and Oscillation}
In the $90^\circ$ flux emergence case, a prominence is formed when the emerged loops get stretched and reconnected between two legs. Due to continuous reconnection, a filament channel evolves over time and efficient plasma injection leads to the formation of larger and higher-altitude filaments. The stratified solar atmosphere plays a critical role in this process, as the temperature and density gradients between the chromosphere, TR, and corona facilitate the upward transport and confinement of dense plasma. The observed filament diameters and heights in our simulations, increasing from 6 Mm to 10 Mm and from 22 Mm to 40 Mm respectively, align with the expected scaling of filaments formed through reconnection processes. These results are consistent with previous studies, which have highlighted the role of magnetic reconnection in creating filament channels \citep[e.g.,][]{LiH2022}. Furthermore, our results are supported by the observations of \citet{Okamoto2010} and \citet{LiX2012}. \citet{Okamoto2010} observed the rising of the cool column due to the emergence of a helical flux rope that undergoes reconnection with lower coronal fields and carries material into the coronal cavity. \citet{LiX2012} also demonstrate the clear movement of material from the leg of a filament into its cavity along a very fine channel. Our simulation is the first to clearly demonstrate flux-emergence-fed injection leading to filament formation, and shows the complex interplay between magnetic topology, oscillatory reconnection, and filament periodicity.

Once formed, filaments in our simulations exhibit oscillatory and rotational behavior. These motions are triggered by the impulsive reconnection events that inject mass and energy into the filament channel. The oscillations have a period of approximately 10 minutes, with the amplitude initially large and gradually decreasing over time due to damping, consistent with previously decaying filament oscillations \citep[e.g.,][]{Tripathi2009, Zhang2024}. The periodic heating and compression at the filament's leading edge, caused by rebound shocks from upward plasma motions, contribute to the oscillatory and rotational behavior. These dynamics are further enhanced by instabilities such as RTI and KHI, which introduce turbulence into the filament environment and finger-like structures formed beneath the filament during its ascent, while KHI generates roll-up features at the filament boundaries. Our results align well with observations of filament oscillations \citep{Chen2008, Joshi2016, Smirnova2022}, showing that external disturbances from magnetic reconnection events can induce significant oscillatory motions in filaments and subsequent eruptions. These vertical oscillations, observed in many filaments, especially winking filaments, are common during filament evolution \citep{Shen2014, Arregui2018, Zhou2023}. The decay of oscillation amplitude observed in our simulations is also very common, which according to highly analytical and idealized literature studies, can be attributed to energy dissipation through resonant absorption, wave leakage or non-adiabatic processes \citep{Hershaw2011, Arregui2018}. The rotational motion has been modeled by \citet{Valeriia2023} and also reported in observations \citep{Wang2010, LiX2012}. From the observational movies in \citet{Wang2010} and \citet{LiX2012}, we can clearly see the upward motion of the chromospheric plasma into the cavity and the rotational motion of the cavity or formed prominence. The synthetic AIA images in our simulations are very similar to the clear tornado-like appearance of the filament and cavity in the real observations. Future work will need to consider fully 3D counterparts of our setups, where especially the case with filament formation will deliver new insights into our dynamical, multi-thermal solar atmosphere.

\subsection{H$\alpha$ Synthesis Models}

\begin{figure}[p]
  \centering
   \includegraphics[trim=3cm 0cm 3cm 0cm, clip, width=\textwidth]{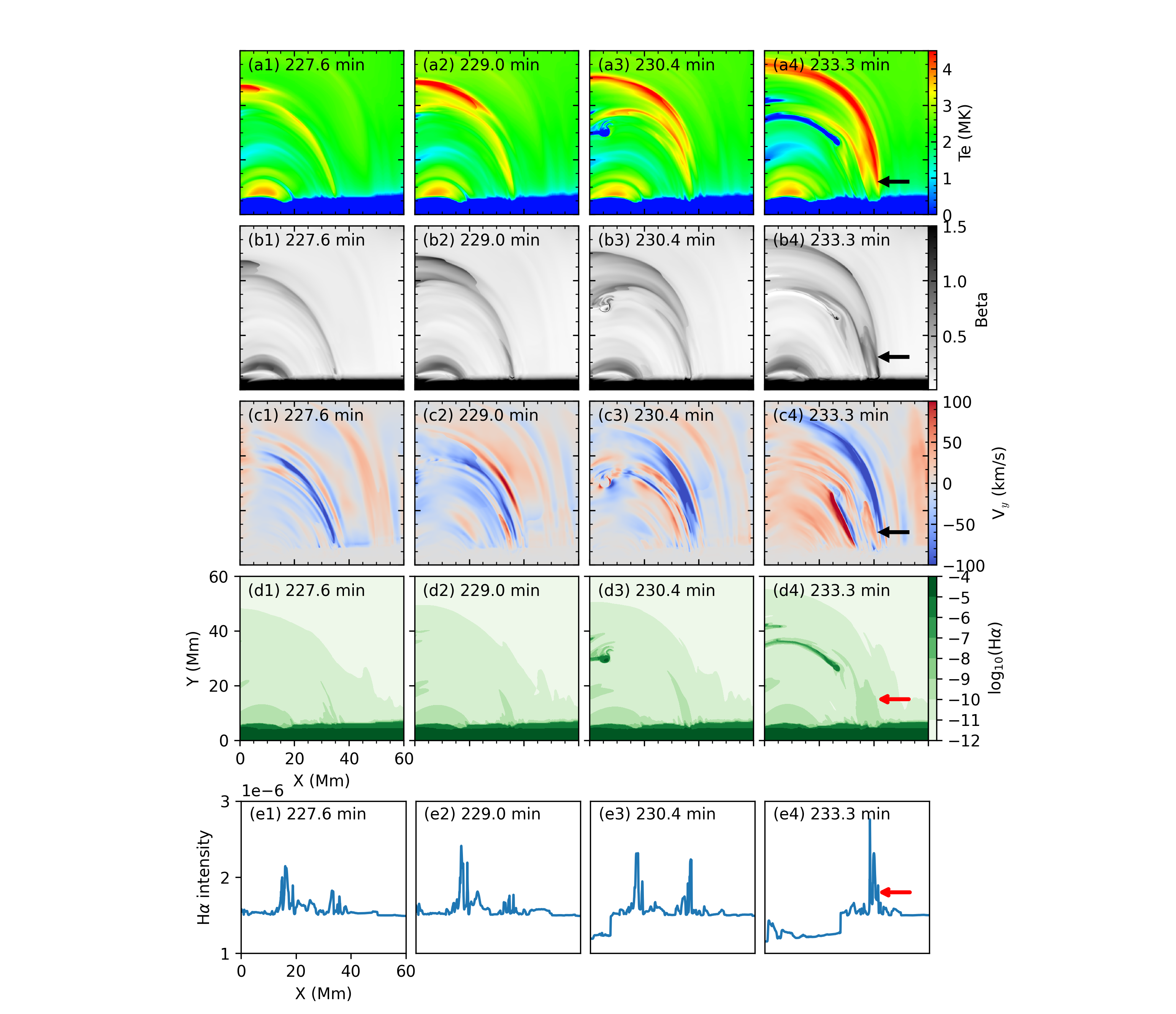}
     \caption{Temperature, plasma beta, $y$-direction velocity, H$\alpha$ synthetic images, and H$\alpha$ intensity curves (from the top view) displaying the response of chromosphere and TR to the hot jet during case B$_1$= 200 G. The arrows in the right panels indicate the backflow. 
     }
        \label{Fig6}
  \end{figure}

\begin{figure}[p]
  \centering
   \includegraphics[trim=0cm 2cm 0cm 1cm, clip, width=\textwidth]{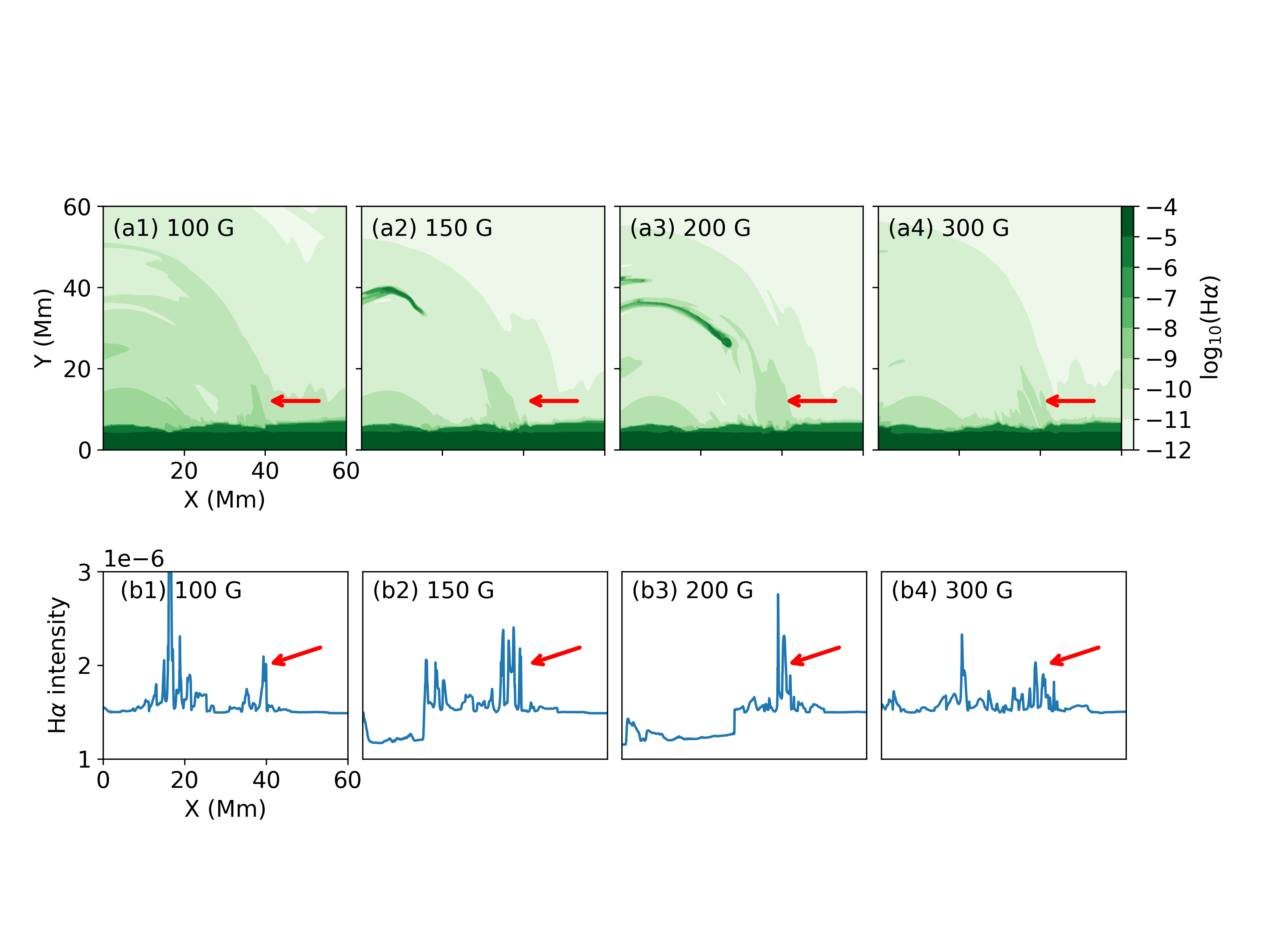}
     \caption{H$\alpha$ synthetic images and intensity curves (from the top view) displaying the response of chromosphere and TR to the hot jet with different emerging field strength. The times for different field strength cases (from left to right: 250.5 min, 237.6 min, 233.3 min, 224.7 min) are chosen exactly when the hot components reach the TR. The red arrows point to the brightenings.  
       }
        \label{Fig7}
  \end{figure}

\begin{figure}[p]
  \centering
  \includegraphics[trim=2cm 1cm 2cm 0cm, clip, width= \textwidth]{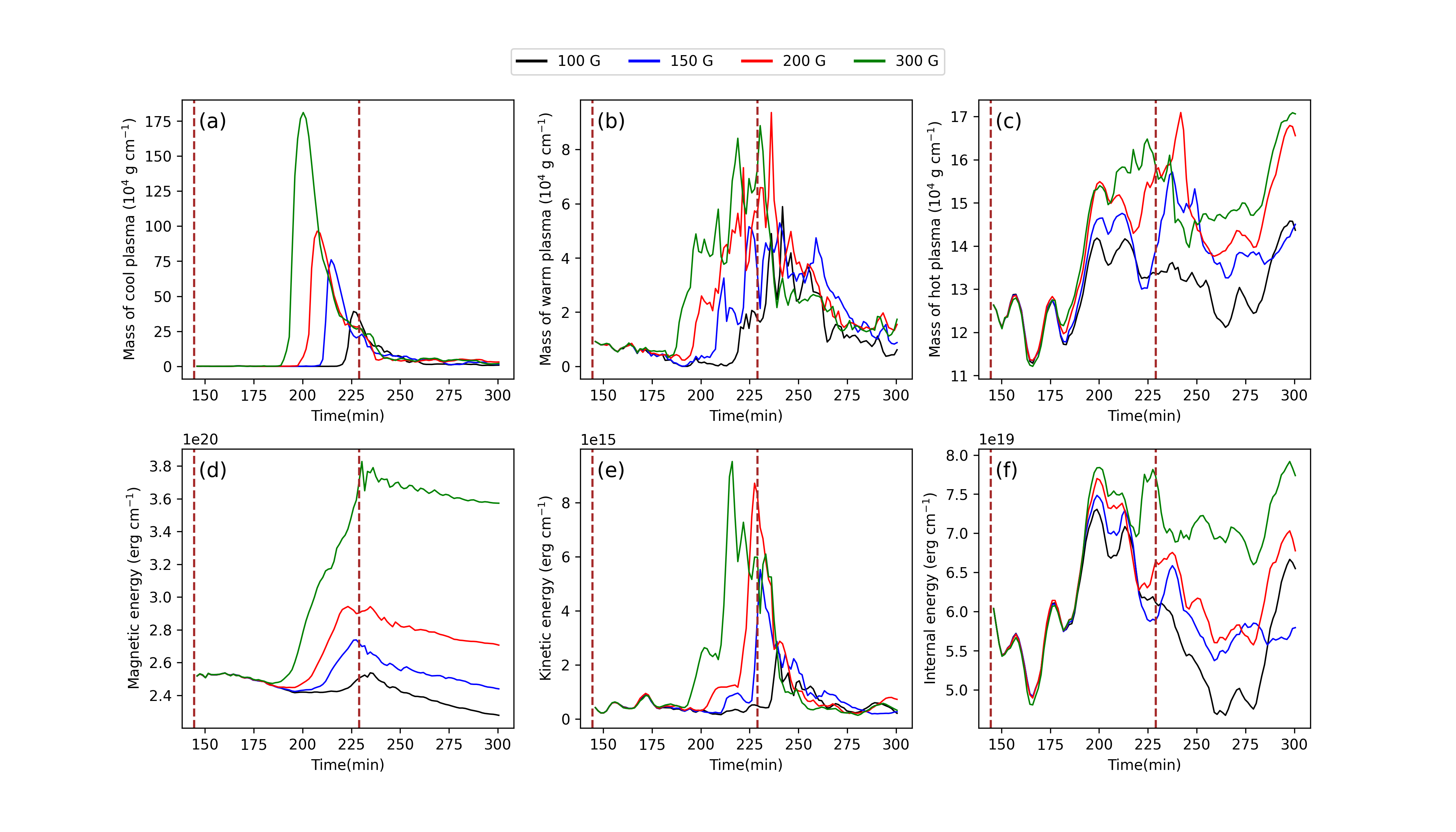}
     \caption{Time evolution of (a) the mass of cool plasma, (b) the mass of warm plasma, (c) the mass of hot plasma, (d) magnetic energy, (e) kinetic energy and (f) internal energy in the corona ($y \ge$ 7 Mm) with different emerging field strength. The vertical dashed lines indicated the time when the emergence starts and ends.
     }
     \label{Fig8}
  \end{figure}

\begin{figure}[p]
   \centering
   \includegraphics[trim=0cm 3cm 0cm 2cm, clip, width=\textwidth]{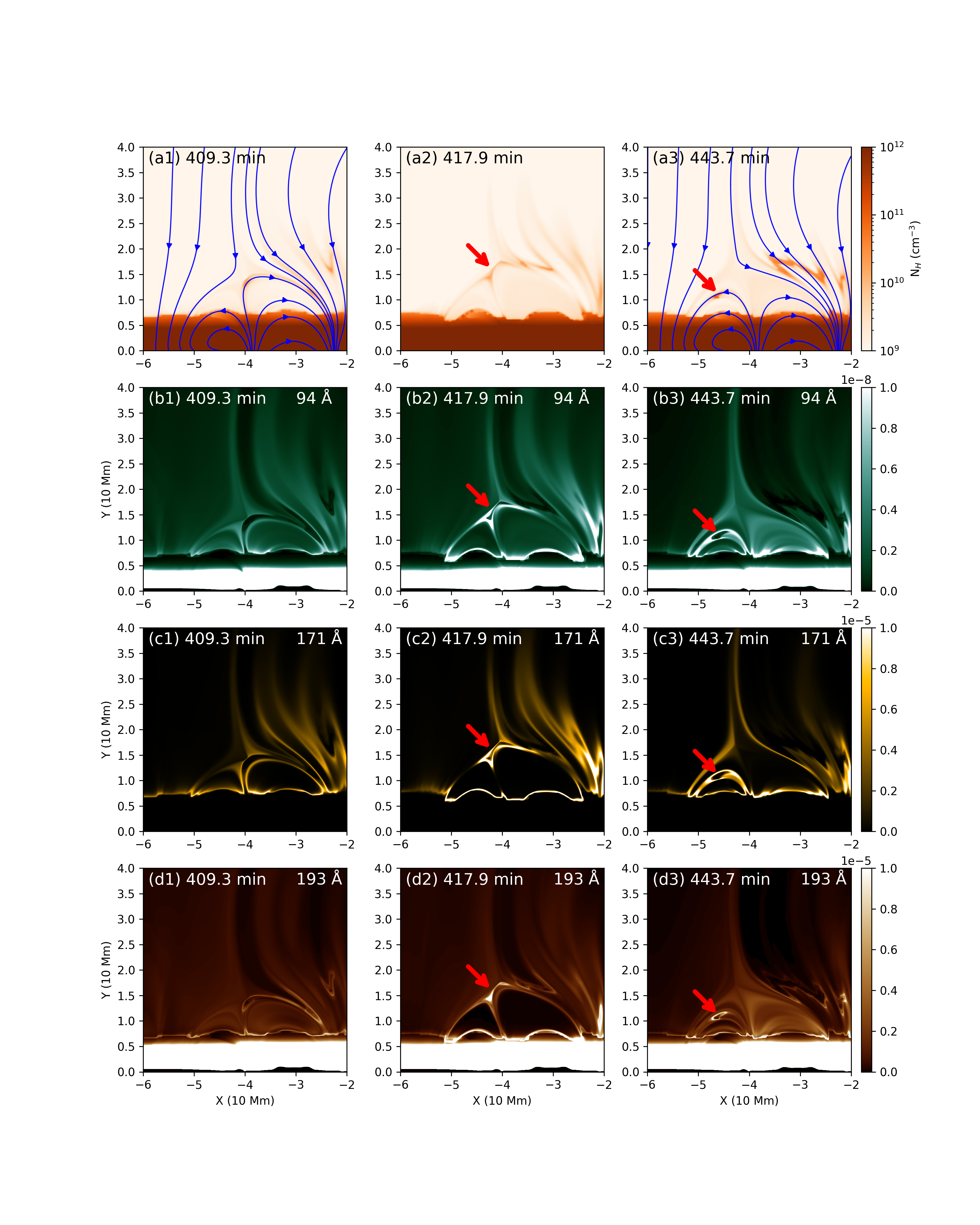}
      \caption{Number density maps and AIA 94 {\AA}, 171 {\AA}, 193 {\AA} synthetic images of magnetic reconnection. The blue curves in panels (a1) and (a3) denote the magnetic field lines. The red arrows in the middle-column panels indicate the location of the current sheet, while those in the rightmost-column panels denote the condensation in the newly formed magnetic loop.}
      \label{Fig9}
   \end{figure}

\begin{figure}[p]
  \centering
  \includegraphics[width=18cm]{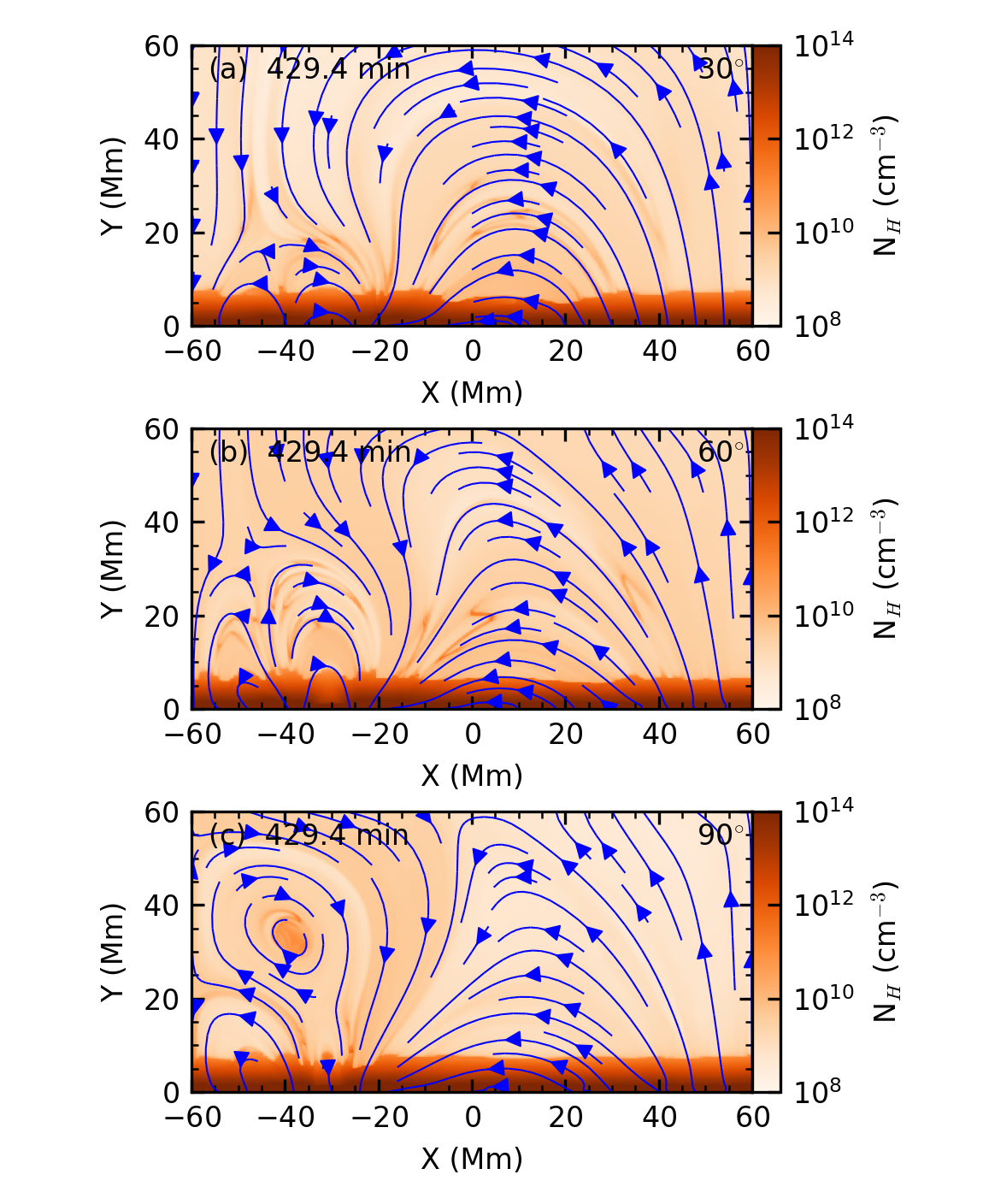}
     \caption{Number density maps showing the eruption caused by the emerging field with relative angle of (a) 30$^{\circ}$, (b) 60$^{\circ}$ and (c) 90$^{\circ}$. The blue curves display the magnetic field lines. An animation of this figure (Movie2.mov) is available online.
     }
     \label{Fig10}
  \end{figure}

\begin{figure}[p]
  \centering
  \includegraphics[width=18cm]{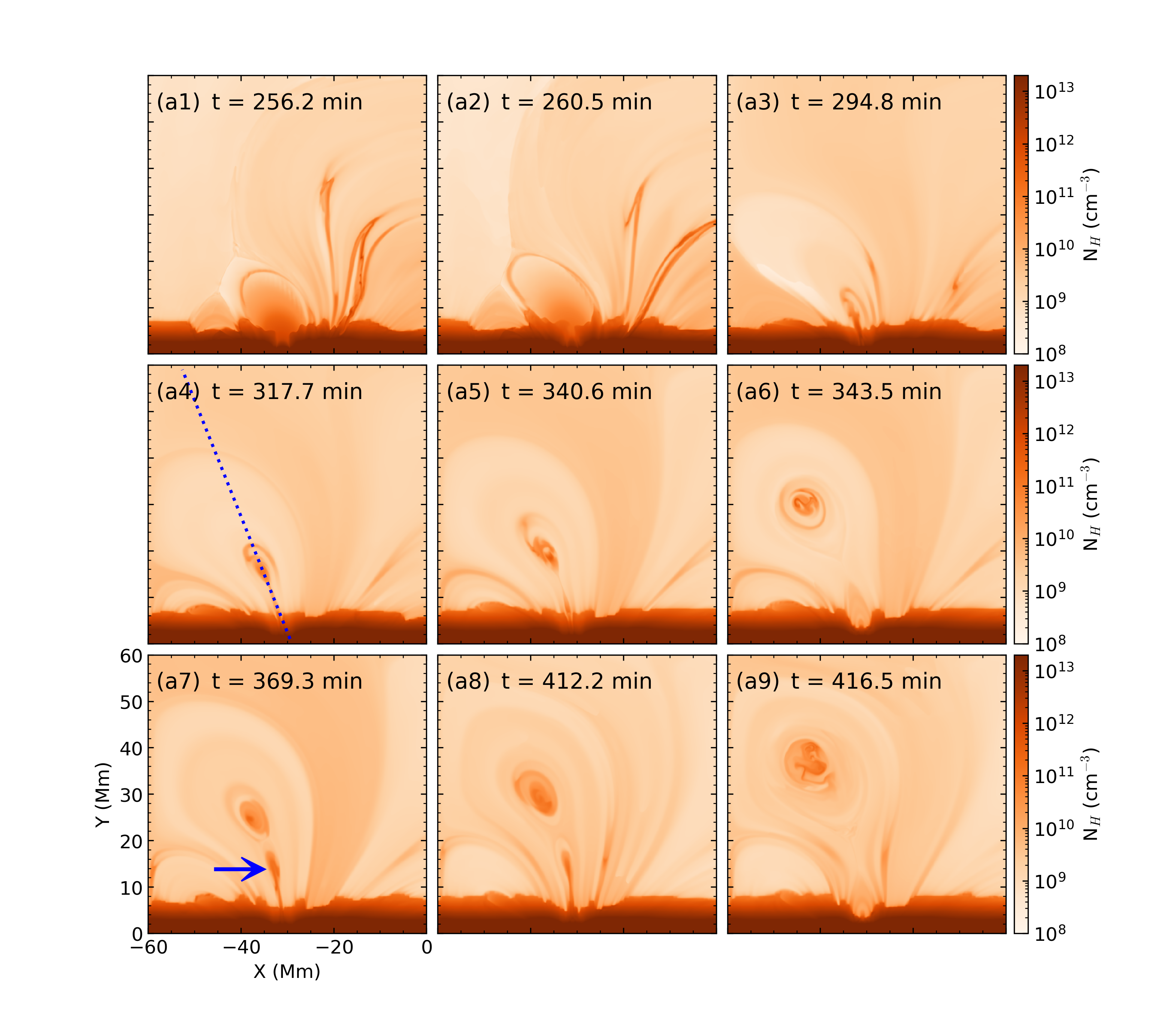}
     \caption{Number density maps showing formation of prominence when the relative angle is 90$^{\circ}$. The dashed line in panel (a5) denotes the slice position for Fig. \ref{Fig13}. The blue arrow in panel (a7) indicates the injection of chromospheric plasma into the filament channel. 
     }
    \label{Fig11}
  \end{figure}

\begin{figure}[p]
  \centering
  \includegraphics[trim=0cm 6cm 0cm 7cm, clip,width=\textwidth]{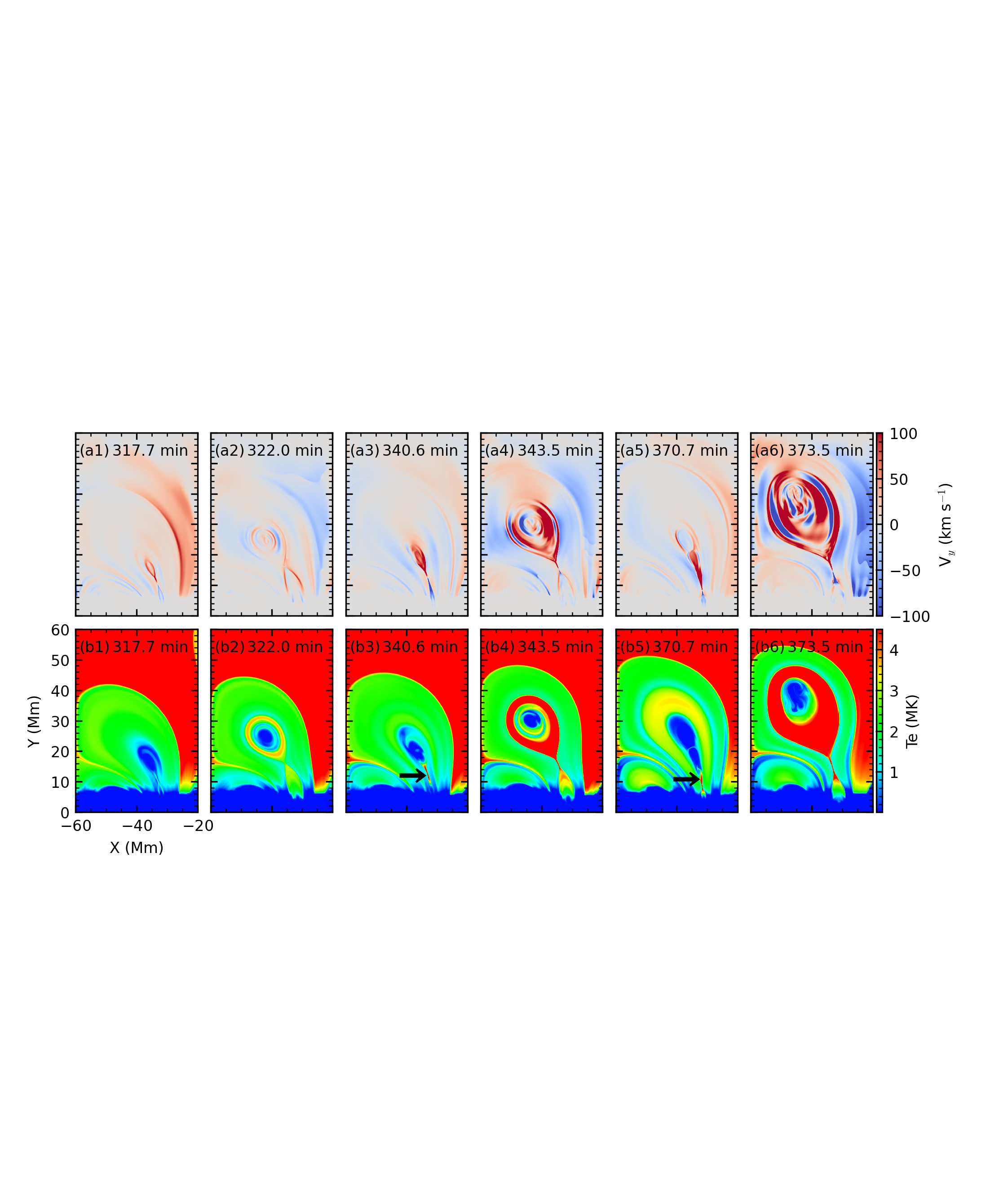}
     \caption{Velocity and temperature maps showing formation of prominence when the relative angle is 90$^{\circ}$. The black arrows point to the hot current sheet.}
      \label{Fig12}
\end{figure}

\begin{figure}[p]
   \centering
  \includegraphics[trim=0cm 2.5cm 0cm 2cm, clip,width=\textwidth]{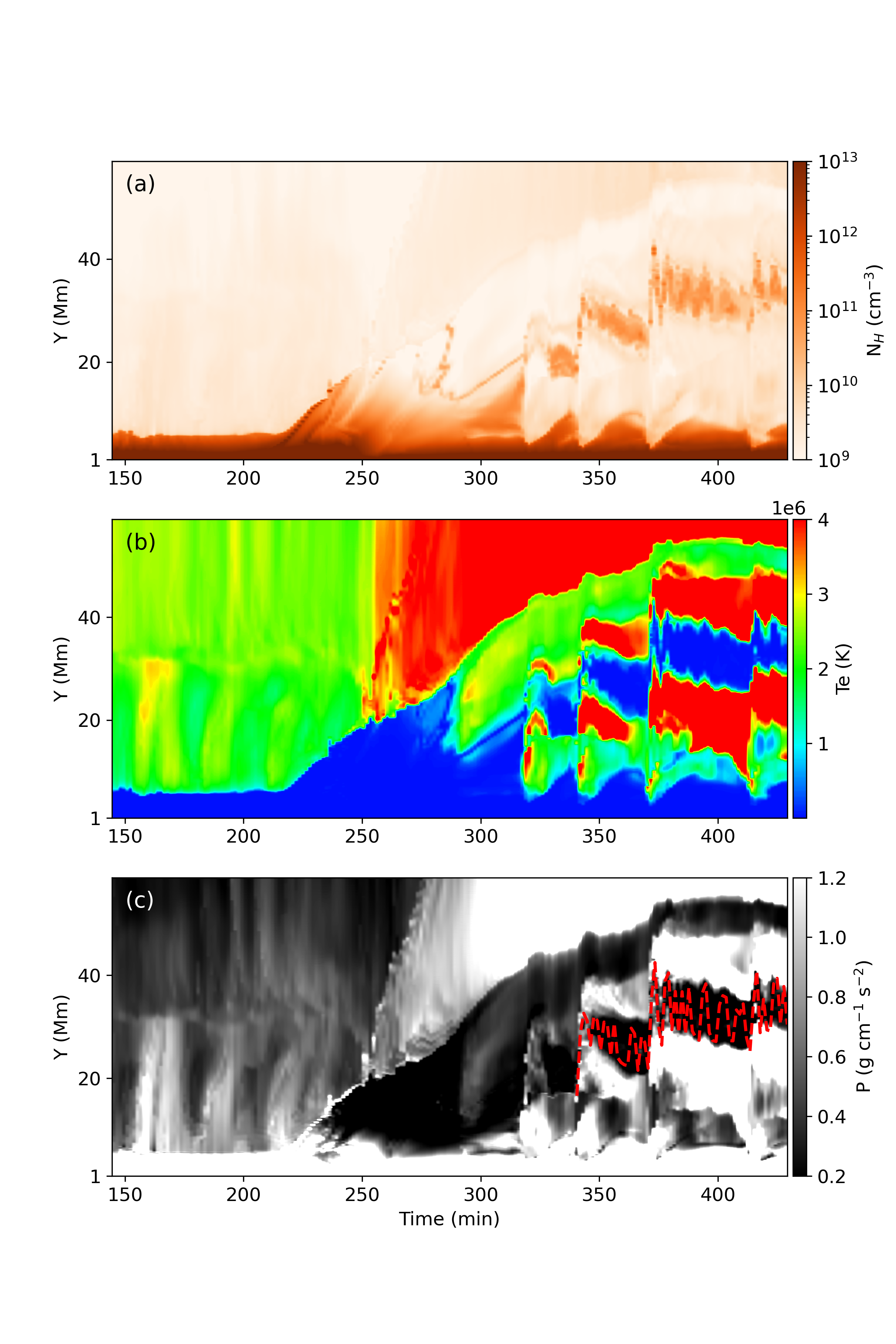}
      \caption{Time evolution of the density, temperature and thermal pressure along the slice in Fig.\ref{Fig11} when when the relative angle is 90$^{\circ}$. The red dashed curves in panel (c) show the core of the prominence where the number density is the highest.
      }
   \label{Fig13}
   \end{figure}

\begin{figure}[p]
   \centering
  \includegraphics[trim=1cm 2cm 0cm 1cm, clip,width=\textwidth]{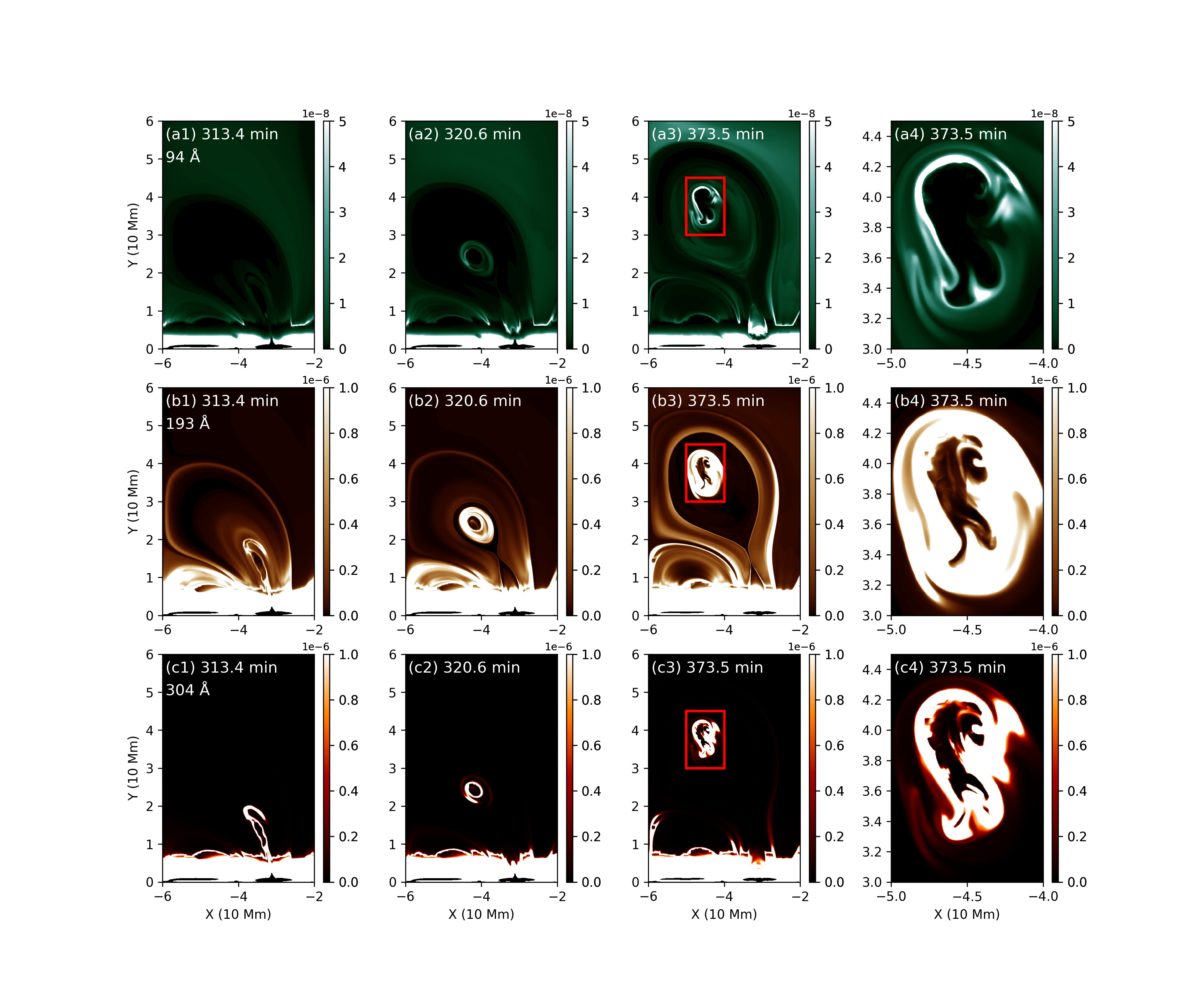}
      \caption{AIA 94 {\AA}, 193 {\AA} and 304 {\AA} synthetic images of the prominence formation when the relative angle is 90$^{\circ}$. The red squares in panels (a3)$-$(c3) indicate the field-of-view of panels (a4)$-$(c4). An animation associated with this figure (Movie3.mov) is available online.}
         \label{Fig14}
   \end{figure}

To interpret the chromospheric response and filamentary features in our simulations, we synthesized H$\alpha$ emission using the semi-empirical approximation proposed by \citet{Heinzel2015}. This method estimates the hydrogen $n=2$ level population ($n_2$) based on local plasma parameters, offering a computationally efficient proxy for optical depth in H$\alpha$. While this approach has been validated for cool and dense prominence or filament conditions, its extension to hotter and more tenuous regimes such as the corona or lower chromosphere must be treated with caution. To minimize these limitations, we focus our interpretation on cool jet material, prominence condensations, and the upper chromosphere and transition region (above $\sim$4 Mm), where the approximation has been shown to yield semi-quantitative agreement with full non-local thermodynamic equilibrium (NLTE) models. Emission from hot coronal loops in Fig. \ref{Fig6} and Fig. \ref{Fig7} is weak and interpreted as a by-product of the approximation rather than a physically meaningful signal. The enhanced H$\alpha$ emission denoted in the figures arises from compression of the transition region and upper chromosphere caused by jet material impacting the footpoints. In future work, we plan to incorporate more advanced NLTE radiative transfer tools to improve the accuracy of H$\alpha$ synthesis and extend the diagnostic capability to a broader range of atmospheric conditions. Examples of such multi-dimensional NLTE synthesis on time-evolving MPI-AMRVAC simulations like ours have been pioneered in \cite{Osborne2025}.

%----------------------------------------------------------- 
  
%----------------------------------------------------------- 
  
  %----------------------------------------------------------- 
  
  %----------------------------------------------------------- 

%----------------------------------------------------------- 
  
%----------------------------------------------------------- 
    
   %----------------------------------------------------------- 
 
%----------------------------------------------------------- 

%----------------------------------------------------------- 

%----------------------------------------------------------- 

%----------------------------------------------------------- 
  
%-----------------------------------------------------------
   
%-----------------------------------------------------------
   
%----------------------------------------------------------- 

\begin{acknowledgements}
We thank the referee for valuable suggestions. This work is done under the support of the European Research Council (ERC) under the European Unions Horizon 2020 research and innovation program (grant agreement No. 833251 PROMINENT ERC-ADG 2018). This work is further supported by an FWO project G0B9923N Helioskill and by the KU Leuven project C16/24/010 UnderRadioSun. YZ acknowledges funding from Research Foundation – Flanders FWO under project number 1256423N. The computational resources and services used in this work were provided by the VSC (Flemish Supercomputer Center), funded by the Research Foundation Flanders (FWO) and the Flemish Government - department EWI. This work uses the open source software \href{https://www.python.org/}{Python} and \href{https://yt-project.org/}{yt} for visualizations. 
\end{acknowledgements}

\vspace{5mm}

\bibliography{manuscript.bib}
\bibliographystyle{manuscript}

\end{document}